# EDUARDO CUNHA DE ALMEIDA

# ESTUDO DE VIABILIDADE DE UMA PLATAFORMA DE BAIXO CUSTO PARA DATA WAREHOUSE

**CURITIBA**

**2004**

# EDUARDO CUNHA DE ALMEIDA

# ESTUDO DE VIABILIDADE DE UMA PLATAFORMA DE BAIXO CUSTO PARA DATA WAREHOUSE

Dissertação apresentada como requisito à obtenção do grau de Mestre. Curso de Pós-Graduação em Informática, Setor de Ciências Exatas, Universidade Federal do Paraná,

Orientador: Prof. Dr. Marcos Sfair Sunye

# CURITIBA
# 2004

Dedico esta dissertação à minha mãe Rosana, minha avó Maria Aparecida e minha futura esposa Siloé que me apoiaram durante todo o tempo em que estive desenvolvendo este trabalho.

Também dedico ao meu "mestre" e orientador Marcos Sunye por sua competência e atenção.

## AGRADECIMENTOS

# SUMÁRIO

## LISTA DE FIGURAS

**LISTA DE QUADROS**

## LISTA DE SIGLAS

| | | |
|---|---|---|
| AS | - | APPLICATION SERVER |
| DBA | - | ADMINISTRADOR DE BANCO DE DADOS |
| DW | - | DATA WAREHOUSE |
| DML | - | LINGUAGEM DE MANIPULAÇÃO DE DADOS |
| DSS | - | DECISION SUPPORT SYSTEMS |
| EIS | - | EXECUTIVE INFORMATION SYSTEMS |
| E/S | - | ENTRADA E SAÍDA DE DADOS |
| ETL | - | EXTRACT, TRANSFORM AND LOAD |
| GDM | - | GENERIC DATA MODEL |
| ODS | - | OPERATIONAL DATA STORAGE |
| OLAP | - | ON-LINE ANALYTICAL PROCESSING |
| OLTP | - | ON-LINE TRANSACTION PROCESSING |
| OSDL | - | OPEN SOURCE DEVELOPMENT LABS |
| SF | - | FATOR DE ESCALA |
| SGBD | - | SISTEMA GERENCIADOR DE BANCO DE DADOS |
| SO | - | SISTEMA OPERACIONAL |
| SQL | - | STRUCTURED QUERY LANGUAGE |
| TI | - | TECNOLOGIA DA INFORMAÇÃO |
| TPC | - | TRANSACTION PROCESSING PERFORMANCE COUNCIL |

# RESUMO

O mundo empresarial necessita cada vez mais de instrumentos que melhorem a tomada de decisões diante do mercado competitivo.  Porém, nem todas as empresas dispõem de recursos para a aquisição dos sistemas existentes e consolidados que o mercado oferece, face o seu alto custo. Esta dificuldade pode ser estendida a outros segmentos, como o governo e universidades, que também necessitam de dinamismo nas tomadas de decisões.

Este trabalho é dedicado ao estudo da viabilidade de uma plataforma de baixo custo para *data warehouse* capaz de atender a esta clientela. Consideramos como plataforma de baixo custo a utilização de software de código aberto PostgreSQL e GNU/Linux.

As características do PostgreSQL são brevemente descritas e apresentadas sugestões de implementações que podem aumentar o desempenho deste SGBD em ambientes de *data warehousing*.

Para verificar a viabilidade desta plataforma em um ambiente de *data warehousing*, executamos *benchmarks* que são medições do desempenho de um sistema sob uma carga de trabalho. Neste trabalho foram utilizados os *benchmarks* TPC-H e DBT3 que simulam a carga de trabalho de um *data warehousing*.

Estes *benchmarks* cobrem um ambiente multiusuario com consultas que realizam operações complexas, como por exemplo, agregações, sub-consultas aninhadas, múltiplas junções, sub-consultas dentro da clausula FROM, entre outras.

Com os resultados aferidos foi possível apontar as dificuldades que o SGBD PostgreSQL teve na execução do TPC-H, motivo pelo qual executamos o DBT3 demonstrando a inviabilidade de utilizar o PostgreSQL versão 7.x como SGBD para *data warehouse*.

Finalmente, em face dos resultados deste estudo, são sugeridas implementações para que este SGBD possa ser utilizado sem restrições em um projeto de *data warehouse*.

# ABSTRACT

Often corporations need tools that increase the power of decisions in a competitive market. Facing that, several companies does not have resources to buy the commercial systems because of the high costs. This problem can be extended to other segments like government and universities.

This work is dedicated to a feasibility study of a low cost platform to data warehouse to supply these customers. We consider as a low cost platform the use of open source software like DBMS PostgreSQL and GNU/Linux operational system.

The PostgreSQL's features are briefly presented and suggestions of implementations to increase data warehouse performance in this DBMS are pointed.

We verify the feasibility of a data warehouse on this platform by executing benchmarks that serves as a point of reference. In this work we used TPC-H and DBT3 benchmarks that simulate a data warehouse workload.

These benchmarks simulate multi-user environment and run complex queries, which executes: aggregations, nested sub queries, multi joins, in-line views and more.

Considering the results we were able to highlight the PostgreSQL's problems in the TPC-H execution, these problems were the main reason to execute the DBT3 benchmark and the reason to invalidate the use of PostgreSQL version 7.x as a data warehouse DBMS.

Finally we make suggestions of implementations to this DBMS becomes available without reservations in data warehouse projects.

# 1 INTRODUÇÃO

Com a crescente demanda por informação nas empresas, tendo em vista a grande competitividade no mercado, a utilização de sistemas que ajudam na tomada de decisão cresce a cada dia e tem se mostrado como um importante diferencial competitivo.

Esta demanda está sendo suprida pela utilização de sistemas de apoio à decisão ou *Decision Support System* (DSS) através de ferramentas de *On-line Analytical Processing* (OLAP) e *Data Mining*, ferramentas estas que fazem acesso a um Data Warehouse (DW).

As ferramentas de OLAP são utilizadas para a criação de relatórios gerencias facilitando a formatação multidimensional dos dados. Estes relatórios podem ser disponibilizados em ambiente *web,* e então, acessadas pelos tomadores de decisão numa intranet, por exemplo.

As ferramentas de *data mining* utilizam técnicas estatísticas com a finalidade de encontrar correlações que ajudem nas decisões estratégicas. Estas ferramentas são normalmente utilizadas pelos analistas para então gerar informações relevantes que possam ser acessadas pelas ferramentas OLAP.

O d*ata warehouse* é um banco de dados que integra num único repositório os dados necessários ao suporte à tomada de decisão e que fornece os dados para as ferramentas de OLAP e *data mining*.

Estes dados são carregados através de programas que acessam outros bancos de dados, extraem os dados e os inserem na base d*ata warehouse*.

ELMASRI e NAVATHE (2000, p. 842) caracterizam "*data warehouse* como provedor de acesso a dados para análises complexas, descoberta de conhecimento, e suporte a decisão".

Segundo COREY, et al (2001, p. 2), "um *data warehouse* é um banco de dados reunido a partir de muitos sistemas e destinados a oferecer suporte à produção de relatórios gerenciais e à tomada de decisão", como mostrado na FIGURA 1.

A partir destes autores é possível perceber a importância do d*ata warehouse* para o negócio das empresas.

KAVALCO (2001) cita que o processo de criação e manutenção deste tipo de ambiente reúne uma diversidade de passos, como mostrado na FIGURA 1. Podemos destacar a integração dos modelos origem em um modelo consolidado, a transferência dos dados dos aplicativos transacionais de origem para o ambiente integrado, a qualificação, agregação e disponibilização no ambiente de data warehouse e a visualização dos dados disponibilizados.

FIGURA 1 – VISÃO GERAL DE UM *DATA WAREHOUSE*

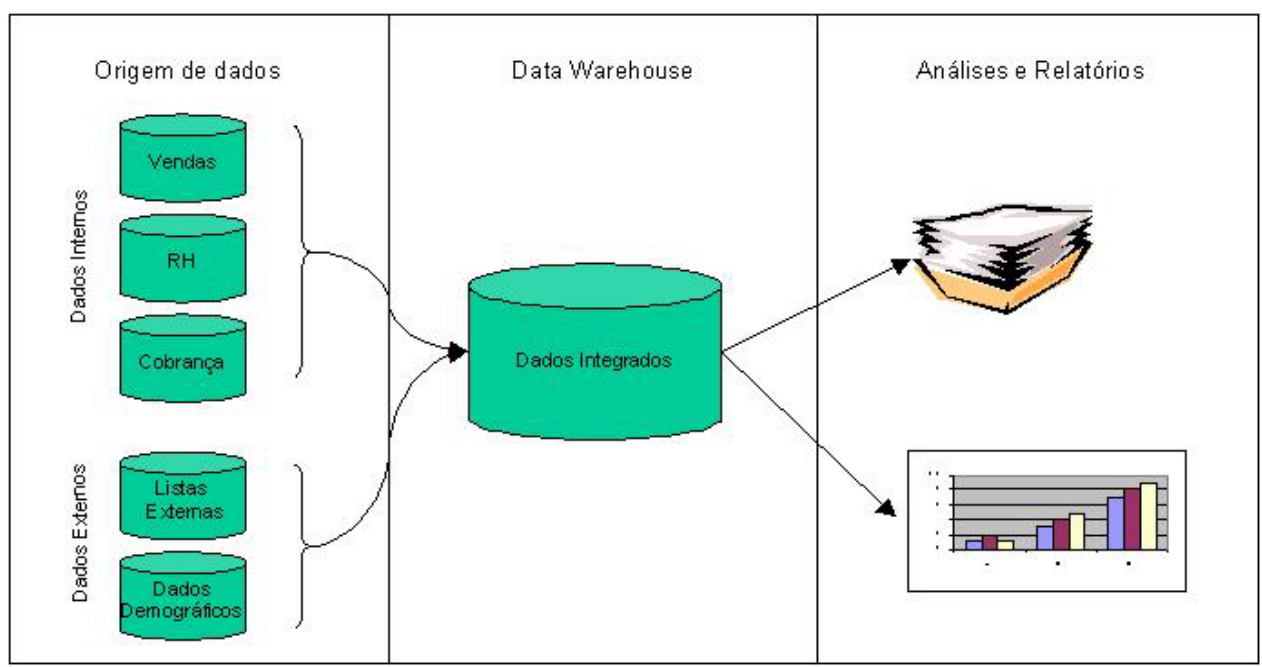

A metodologia utilizada na construção do *data warehouse* garante um grande ganho de desempenho em comparação com a metodologia utilizada em ambientes chamados transacionais ou *On-Line Transaction Processing* (OLTP), nos quais são utilizadas as formas normais de *Boyce/Codd*.

Ambientes de *data warehousing*, pelas dimensões de tamanho e importância estratégica, são extremamente caros, sendo este fator a barreira para o desenvolvimento em universidades e empresas com pouco capital.

Parece promissor, portanto, o estudo de viabilidade de uma plataforma de baixo custo para d*ata warehouse* visando possibilitar o desenvolvimento deste ambiente em organizações como as citadas acima.

Neste estudo utilizamos software conhecido como "*open source*" ou software de código aberto. Em particular, o sistema operacional GNU/Linux e o SGBD PostgreSQL, bem como equipamentos de baixo custo, buscando obter um desempenho viável, visando motivar o desenvolvimento e melhoria de softwares de código aberto para os diversos componentes do ambiente *data warehouse.*

Na medição de desempenho utilizamos as metodologias desenvolvidas pelo Open Source Development Lab (OSDL) e pelo Transaction Processing Performance Council (TPC) nas escalas de 1 GB e 100 GB e explicaremos o motivo da utilização de duas metodologias e de duas escalas.

O OSDL é uma organização sem fins lucrativos que fornece o "estado da arte" em computação e ambientes de teste com a finalidade de acelerar o crescimento e a adoção do sistema operacional GNU/Linux nas empresas. Estes ambientes estão localizados nos EUA e Japão e estão disponíveis para desenvolvedores de todo o mundo. Esta organização é mantida por um consórcio global de empresas entre elas IBM, HP, CA, Intel e NEC.

O TPC é uma corporação sem fins lucrativos fundado para definir processamento de transações e *benchmarks* de banco de dados.

O propósito do *benchmark* TPC é prover dados de desempenho relevantes e objetivos para a indústria.

Hoje as grandes empresas que possuem seus produtos envolvidos no *benchmark*, sejam estes produtos software ou hardware, são membros do TPC como: Oracle, IBM, NCR Teradata, Microsoft, Sybase, Sun, HP e outras.

As metodologias utilizadas como referências para este estudo foram a OSDL DBT3 e a TPC-H, pois medem o desempenho em sistemas de apoio à decisão que se beneficiam do *data warehouse* como fonte de dados.

Neste trabalho descrevemos detalhadamente a metodologia do TPC, pois a OSDL DBT3 utiliza toda a metodologia TPC-H com algumas modificações. Foram apresentadas as diferenças entre as metodologias na mesma seção que será descrita a TPC-H.

O trabalho está dividido em nove capítulos, sendo que o capítulo dois descreve o objetivo do estudo; o capítulo três relata uma revisão da literatura de trabalhos correlatos em *benchmark* de banco de dados; o capítulo quatro descreve os conceitos de *data warehouse* e tópicos de desempenho em SGBD, considerando o desenvolvimento de um *data warehouse* e buscando indicar características para melhoria do SGBD; o capítulo cinco descreve a história e as características do PostgreSQL, bem como uma comparação de valores de SGBDs; no capítulo seis as metodologias TPC-H e DBT3 são analisadas; no capítulo sete são descritas a configuração e as implementações para a realização dos testes de desempenho; no capítulo oito os resultados são detalhados e a viabilidade da plataforma de *data warehouse* de baixo custo analisada e no capítulo nove é apresentada a conclusão.

## 2 OBJETIVO

O objetivo deste trabalho é a realização de um estudo de viabilidade de uma plataforma de baixo custo para *data warehouse*.

Através deste estudo esperamos demonstrar que são possíveis e promissores esforços no desenvolvimento do software envolvidos visando suporte a ambientes de *data warehousing*.

Espera-se, portanto, gerar uma alternativa para organizações que necessitam de um *data warehouse* e não possuem capital necessário para adquirir os atuais sistemas, ou buscam esta implementação aceitando a relação entre desempenho e custo.

A medição de desempenho será atestada através de metodologias de *benchmark*. Portanto, será aplicada uma metodologia para avaliação de desempenho de uma estrutura com máquina e SGBD de baixo custo em comparação com estruturas proprietárias publicados pelo TPC.

Consideramos como plataforma de baixo custo a utilização de software de "código aberto", sendo eles SGBD PostgreSQL, sistema operacional GNU/Linux, Java e equipamento de baixo custo.

Usando o modelo de código aberto esperamos, inclusive, fomentar o interesse no desenvolvimento de softwares para os outros componentes que fazem parte do ambiente de *data warehousing*.

Apesar deste estudo estar focado na viabilidade do desempenho do SGBD e do sistema operacional em equipamento de baixo custo, também são citados estudos realizados para melhoria de SGBDs.

## 3 REVISÃO DE LITERATURA

Alguns trabalhos foram desenvolvidos envolvendo *benchmarks* de SGBDs, propondo metodologias e realizando medições de desempenho.

Estes trabalhos buscaram, primordialmente, SGBDs para operações transacionais. Com o surgimento do *data warehouse* na década de 90 diversas empresas apresentaram suas soluções, como Oracle, IBM e Sybase, surgindo então a necessidade de realizar *benchmarks* voltados para avaliar e comparar desempenho, bem como o custo dos componentes envolvidos no projeto.

O TPC aparece como o principal organismo realizador de *benchmarks* com o propósito de prover dados de desempenho relevantes e objetivos para a indústria. Para alcançar esta proposta a especificação do TPC requer que o teste de *benchmark* seja implementado com sistemas, produtos, tecnologias e preços que:

- São geralmente disponíveis aos usuários;
- São relevantes ao segmento de mercado;
- Seja plausível a implementação por um número significativo de usuários. TPC (2003).

O *benchmark* proposto pelo TPC para d*ata warehouse* é conhecido como TPC-H.

O TPC-H define um banco de dados sintético e possui um gerador de dados para esta base. Ele também descreve as consultas que deverão acessar o SGBD em questão, simulando um ambiente real.

As métricas utilizadas para aferir os resultados são produtividade, preço-desempenho e disponibilidade do sistema.

As medições realizadas pelo TPC utilizam SGBDs, sistemas operacionais e hardware proprietários como, por exemplo: SGBD Oracle em máquina HP com Unix Tru64; SGBD Sybase em máquina Sun com Unix Solaris ou SGBD SQL Server em máquina Dell com Windows 2000.

A metodologia empregada pelo TPC-H é analisada em uma seção própria.

O trabalho apresentado por KIM, W. e GARZA, J. F. (1995), descreve alguns requisitos para a realização de *benchmarks*. Este estudo foi desenvolvido para medições em SGBD orientado a objetos.

Neste estudo estão descritas operações essenciais em SGBDs orientado a objetos consideradas como requisitos para que um SGBD seja elegível a este *benchmark,* bem como passos gerais para implementar tal *benchmark*.

Apesar deste *benchmark* somente envolver SGBD orientados a objetos, alguns requisitos importantes são detalhados.

Segundo KIM e GARZA (1995, p. 213) "se um *benchmark* de banco de dados é pequeno, ou o conjunto de operações medidas sobre o banco de dados é somente uma fração de um conjunto de operações compreensivas e que precisam ser medidas, então o *benchmark* tem pouco significado".

Outro requisito importante é a necessidade de ser um ambiente multiusuário, pois um SGBD que não permite este tipo de ambiente, não é considerado útil KIM e GARZA (1995, p. 213).

BITTON, D. et al (1983) descrevem uma metodologia de *benchmark* conhecida como "Wisconsin Benchmark". Este trabalho foi apresentado como o primeiro *benchmark* para SGBDs e foi realizado em ambiente monousuário, pois a finalidade era avaliar os impactos de cada operação no sistema.

O "Wisconsin Benchmark" utiliza um banco de dados sintético e as consultas implementadas para os testes visam medir os custos de diferentes tipos de operações. O resultado deste trabalho é descrito em fator de desempenho.

O trabalho de BITTON considera como fator de desempenho o tempo de resposta de uma consulta no SGBD "A" dividido pelo tempo de resposta no SGBD "B".

Nos resultados apresentados, os SGBDs comerciais apresentaram fatores de desempenho maiores se comparado com o SGBD universitário, ou seja, os sistemas comerciais apresentaram melhor desempenho.

BORAL, H. e DEWITT, D. (1984) apresentam uma metodologia de avaliação de desempenho para ambientes multiusuário.

Alguns aspectos são considerados importantes neste trabalho como:

-   Segmentação das consultas em 4 tipos;

- Limite de 21 consultas para o *benchmark*;
- Caracterização do "mix" de consultas para simular um ambiente real;
- Métricas para a avaliação dos resultados.

A segmentação das consultas em 4 tipos é realizada considerando a baixa e a alta taxa de processamento e operação de entrada e saída (E/S) em disco.

O trabalho também descreve que a utilização de 21 consultas para o *benchmark* é suficiente para caracterizar o desempenho de um sistema (o TPC utiliza 22 consultas em sua metodologia).

TURBYFILL, C. et al (1993) desenvolveram uma metodologia de *benchmark* para SGBDs OLTP que se propõe medir todos os tipos de SGBD e não somente grandes sistemas de bancos de dados. O estudo envolve medições considerando os modos monousuário e multiusuário em um banco de dados sintético. As consultas implementadas para o modo monousuário utilizaram as funções básicas dos SGBDs segmentadas de acordo com os diversos índices implementados através de "b-tree" ou "hash". No modo multiusuário foram utilizados vários "*mix*" de consultas, buscando modelar diferentes perfis de carga de trabalho. Nesta medição, são utilizadas desde consultas simples até grandes transações.

O'NEIL, P. (1993) descreve um *benchmark* apresentado como primeiro na avaliação de sistemas DSS, *Marketing Information Systems* e *Direct Marketing*. Neste trabalho são utilizadas consultas dependentes da aplicação, num total de 69. O banco de dados implementado possui somente uma tabela, sendo que operações de junção são realizadas através de *self-joins.* O ambiente em modo multiusuário é simulado através do conjunto de consultas que executam em paralelo. Como neste *benchmark* não rodam transações, é presumida, dada a natureza das aplicações, que as consultas não possuem interferência e, portanto, o conjunto de consultas simula o ambiente multiusuário adequadamente.

A métrica utilizada é Preço em Dólar por Consulta por Segundo ($PRICE/QPS) calculado a partir dos tempos de resposta e dos valores dos componentes do sistema. Foram utilizados dois SGBDs comerciais IBM System/370, o DB2 e o MODEL 204.

QUADRO 1 – COMPARAÇÃO ENTRE METODOLOGIAS

| | Multiusuário | Requer ACID | Tamanho do BD | Tipo do Ambiente | Escala do BD |
|---|---|---|---|---|---|
| TPC-H | Sim | Sim | 8 tabelas | DSS | Sim |
| BITTON | Não | ND | 4 tabelas | OLTP | Não |
| BORAL | Sim | ND | 2 tabelas com 16 copias | OLTP | Não |
| TURBYFILL | Sim | Não | 4 tabelas | OLTP | Sim |
| O´NEIL | Sim | ND | 1 tabela (usa join recursivo) | DSS | Não |

O QUADRO 1 compara as metodologias estudadas demonstrando que a metodologia TPC-H é a mais completa e adequada para a carga de trabalho de um *data warehouse*.

Na comparação entre as metodologias a TPC-H mostra-se como uma compilação de todas, empregando, por exemplo, a quantidade e tipos de consultas descritas por BORAL, H. e DEWITT, D. (1984) ou a métrica utilizada para avaliar o desempenho descrito por O'NEIL, P. (1993).

Em comparação com a metodologia de O'NEIL, que é a outra metodologia para ambientes DSS, a metodologia TCP-H representa melhor o mundo real com um esquema mais elaborado e facilmente encontrado em qualquer empresa.

**4 DATA WAREHOUSE**

4.1 CONCEITO

Os *Data Warehouses* ou Armazéns de Dados surgiram da necessidade de um ambiente integrado, onde fossem armazenados e gerenciados eficientemente os dados produzidos pelas aplicações operacionais.

KAVALCO (2001) coloca um ponto importante na evolução do conceito de um *Data Warehouse* que deixa de ser um sistema de apoio à decisão, ou sistema de informações para executivos, para se tornar um sistema de apoio à corporação.

*Data Warehouse,* portanto, é um banco de dados contendo o histórico de uma organização, onde suas diversas áreas, que necessitem utilizar o d*ata warehouse*, podem planejar suas estratégias através dos seus dados.

Geralmente o d*ata warehouse* é utilizado pelos setores administrativos e pela alta gerência, no desenvolvimento de análises e relatórios. Os dados ali inseridos foram previamente consolidados, sumarizados e depurados para serem utilizados num ambiente orientado a consultas com uma modelagem específica para esta finalidade.

Um ambiente de *data warehousing* pode ser implementado de algumas formas, sendo que, dependendo da escolha do ambiente, os custos podem aumentar.

Podem ser citadas como formas de implementação:

1. *Data Warehouse* Corporativo, onde envolverá muito esforço, dinheiro e comprometimento por parte de todas as equipes envolvidas, mas principalmente de um patrocinador forte. Este patrocinador normalmente é o diretor de Tecnologia de Informação (TI) da empresa;

2. *Data Warehouse* virtual é a construção de visões a partir das bases operacionais. Podem ser utilizadas as visões materializadas para melhorar o desempenho;

3. *Data Marts,* que são pequenos *data warehouses*, geralmente podem ser departamentais ou de qualquer outra forma, dependendo da necessidade. *Data Marts* são menores, mais baratos e mais rápidos de serem implementados do que o um *data warehouse* Corporativo. É considerada uma boa solução para empresas que necessitam deste tipo de sistema, mas que não possuem grandes recursos.

As próximas subseções descrevem brevemente assuntos que influenciam no desempenho do SGBD, bem como o poder de processamento de um ambiente de *data warehouse.*

## 4.2 OTIMIZAÇÃO E DESEMPENHO

O desempenho de um SGBD para *data warehouse* é um dos pontos mais críticos tanto na utilização quanto na justificativa dos altos investimentos que esta tecnologia emprega.

BELL (1988) descreve que o valor de um sistema depende de "quão bem" ele cumpre suas funções provendo um serviço utilizável e efetivo.

KIMBALL, citado por KAVALCO (2001, p 14) comenta que "embora existam grandes avanços sendo feitos em hardware, principalmente na área de processamento paralelo, o futuro do *data warehouse* pertence ao software". Ele lista alguns dos pontos que deverão melhorar significativamente nos próximos anos com base no estudo de KIMBALL divulgado em 1996:

a) otimização das estratégias de execução para consultas de junção em esquema estrela;

b) indexação de tabelas de dimensão, em especial tabelas de muitos milhões de linhas;

c) acesso e indexação da chave composta de grandes tabelas de fatos;

d) extensão da SQL para processar consultas do estilo OLAP;

e) suporte a processamento paralelo;

f) ferramentas para projeto de bancos de dados multidimensionais;

g) ferramentas para extração e administração;

h) ferramentas de consulta para o usuário final.

O desempenho de um ambiente d*ata warehousing* é citado como um dos três grandes desafios a serem enfrentados (...)KIM (2002). Este desempenho começa a ser delineado desde a modelagem, pois nela podemos verificar, por exemplo, quantas junções serão necessárias para a recuperação de informações.

Vários pontos, além da modelagem, devem ser verificados durante todo o projeto e não somente num processo de melhoria de desempenho. Na modelagem já é possível conseguir um rendimento melhor do SGBD dependendo do esquema utilizado.

Outro assunto que impacta no desempenho de um SGBD é a imposição de integridade referencial num esquema de data warehouse. Sistemas Gerenciadores de Banco de Dados (SGBD), como exemplo o Oracle, possuem estatísticas de utilização de tabelas. A implementação da integridade ajuda o SGBD na tomada de decisão de como unir tabelas e também ajuda os desenvolvedores e pessoal de manutenção na garantia da consistência dos dados armazenados no repositório (...) (COREY, et al, 2001, p. 185).

A utilização de índices é outro assunto que deve ser cuidadosamente estudado, pois uma criteriosa implementação pode trazer ótimos resultados para o ambiente como um todo.

Serão descritos nesta seção assuntos que envolvem a otimização e desempenho de um SGBD para *data warehouse*, são eles: *cache* de dados, armazenamento dos dados e índices.

## 4.2.1 *CACHE* DE DADOS

Em um *data warehouse,* grandes volumes de dados são requisitados o tempo todo gerando muito processamento de E/S. Processamento de E/S acarreta em perda de desempenho e portando demora na recuperação de informações para o usuário final.

Segundo AILAMAKI et al (2001, p. 1) a comunicação entre CPU e o armazenamento secundário (E/S) tem sido tradicionalmente reconhecido como o maior gargalo no desempenho de SGBDs. Este gargalo ocorre porque na submissão de uma consulta ao SGBD é verificada, primeiramente, a possibilidade de resolvê-la em memória (*cache*), se não for possível então o disco rígido é acessado (operação de E/S), após este acesso os dados são carregados em memória para então apresentar o resultado ao usuário.

JEUNOT (2000) descreve que operações excessivas de paginação e *swap* também podem ter impacto sobre o desempenho do SGBD e do sistema operacional. A utilização da memória também pode comprometer os percentuais de acerto (*hits*) do banco de dados.

Trabalhos citados por AILAMAKI et al (2001, p.2) afirmam que ocorre 90% de faltas (*misses*) em memória de dados de sistemas DSS.

*Data warehouses* possuem como características consultas chamadas "Ad-hoc", que são consultas não repetitivas. Esta característica significa a pouca reutilização da mesma consulta, então a troca dos dados entre memória e disco rígido é muito intensa. Esta constante troca conseqüentemente gera o alto índice de faltas (*misses*) na *cache*.

A partir deste cenário é possível perceber a importância da otimização de acesso aos dados. A otimização de acesso é implementada através das *caches* de dados ou *buffers* que armazenam as informações recuperadas do disco por algum tempo. Alguns parâmetros devem ser estudados para uma ótima configuração, como: tamanho desta *cache* e também como os objetos ficarão armazenados.

SGBDs utilizam a *cache* para armazenar os dados que mais serão utilizados ou que foram recuperados recentemente, dependendo da estratégia escolhida pelo administrador do banco de dados (DBA). Este tipo de área otimiza o acesso aos dados, pois se estes já estiverem em memória uma operação de E/S pode ser poupada.

Alguns SGBDs, como por exemplo o Oracle, possuem a opção de manter tabelas em memória, ou em disco respeitando uma estratégia de *Last Recent Used* (LRU). JEUNOT (2000) descreve que a determinação dos blocos que serão descarregados da memória é definida por uma lista LRU.

A utilização correta da memória é de grande importância no desempenho do *data warehouse*, pois os dados são recuperados da memória em vez do disco, otimizando o acesso aos dados, mas sempre observando a característica de consultas *ad-hoc* citado anteriormente.

Sendo o dimensionamento desta área um ponto importante na otimização do desempenho, sua configuração deve ser bem cuidadosa para não haver distorção na utilização da memória.

Os problemas de um mau dimensionamento podem ser:

- Excesso de memória alocada, acarretando na falta de memória para outras operações;
- Pouca memória alocada para a *cache*, aumentando os acessos ao disco.

## 4.2.2 ARMAZENAMENTO DOS DADOS

A estratégia de armazenamento dos dados em blocos de informação ou páginas do disco rígido é outro fator que também influencia o desempenho.

Na criação do repositório de um banco de dados OLTP é utilizado normalmente páginas de disco com tamanho variando entre 2Kb e 4Kb. Em um *data warehouse* os valores são maiores, sendo dimensionados a partir de 8Kb.

O tamanho é maior para *data warehouse*, pois os dados de uma *tupla* contêm informações "desnormalizadas". Sendo os dados armazenados numa estrutura "desnormalizada" e utilizando tamanhos maiores de páginas, evita-se a leitura de mais de uma página, fenômeno chamado de encadeamento de linha. O encadeamento de linha acontece quando numa recuperação de informação o SGBD acessa mais de uma página de informação para os valores de uma mesma *tupla*. Portanto a utilização de páginas grandes é indicada para leituras seqüenciais e linhas muito grandes.

JEUNOT (2000) descreve que as páginas maiores melhoram o desempenho de leituras de índice. As páginas maiores podem manter mais entradas de índice em cada página, o que reduz o número de níveis dos índices grandes. Menos níveis de índice significam menos operações de E/S durante a passagem pelas ramificações do índice. Este aspecto se refere somente a bancos de dados de *data warehouse*, pois utilizar páginas grandes em bancos de dados OLTP pode gerar muitas disputas nas páginas de índice durante os processos de atualização.

A FIGURA 2 demonstra a estrutura de uma página de acordo com ERNST et al (1999).

FIGURA 2 – EXEMPLO DE PÁGINA DE DISCO

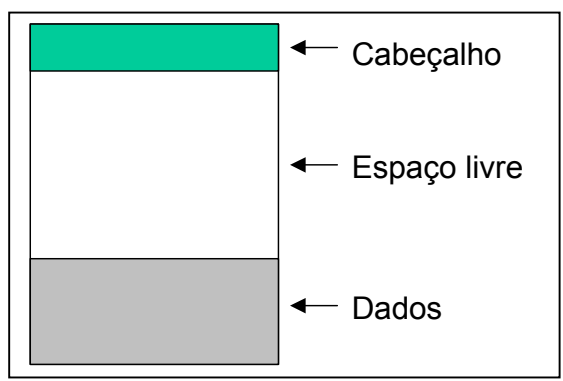

A FIGURA 2 demonstra a página de disco do SGBD Oracle da seguinte forma:

- Cabeçalho da página: O cabeçalho contém o endereço da página de dados, o diretório da tabela, o diretório da linha e os *slots* de transação usados quando as transações fazem alterações em linhas da página. Os cabeçalhos das páginas crescem de cima para baixo;
- Espaço de dados: Os dados de colunas são inseridos na página de baixo para cima;
- Espaço livre: O espaço livre em uma página está localizado no meio, permitindo o crescimento do cabeçalho e do espaço de dados da linha quando necessário.

Apesar de haver algumas modificações nas implementações das estruturas das páginas de dados como a do SGBD Oracle, a estrutura básica segue o conceito de *N-ary Storage Model* (NSN) citado por AILAMAKI et al (2001).

O PostgreSQL também possui a mesma estrutura de blocos como demonstrado na FIGURA 3 descrita por MOMJIAN (2001).

FIGURA 3 – PÁGINA DE DADOS DO POSTGRESQL

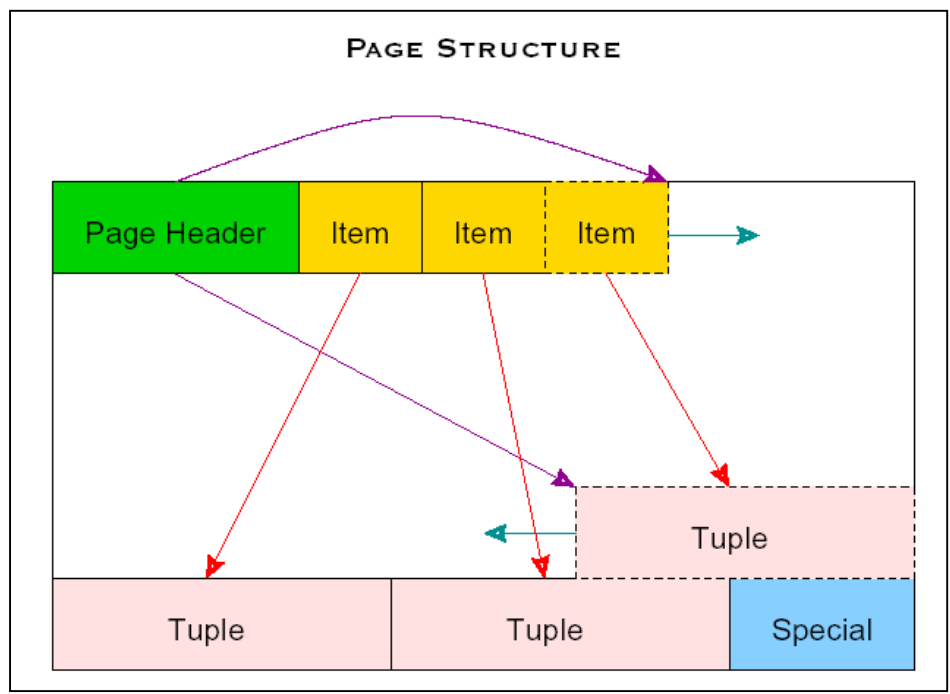

Alguns estudos foram desenvolvidos observando o armazenamento dos dados em bancos de dados, dentre estes AILAMAKI et al (2001) realizaram um trabalho envolvendo bancos de dados para *data warehouse* utilizando a metodologia do TPC.

AILAMAKI et al (2001) apresentam uma estratégia para melhorar o acesso aos dados. Este método chamado de PAX (Partition Attributes Across) demonstrou ganhos de desempenho de 11% à 42% em sistemas DSS utilizando a metodologia TPC-H. A FIGURA 4 ilustra a implementação da estratégia PAX.

FIGURA 4 – EXEMPLO DE PÀGINA DE DADOS PAX

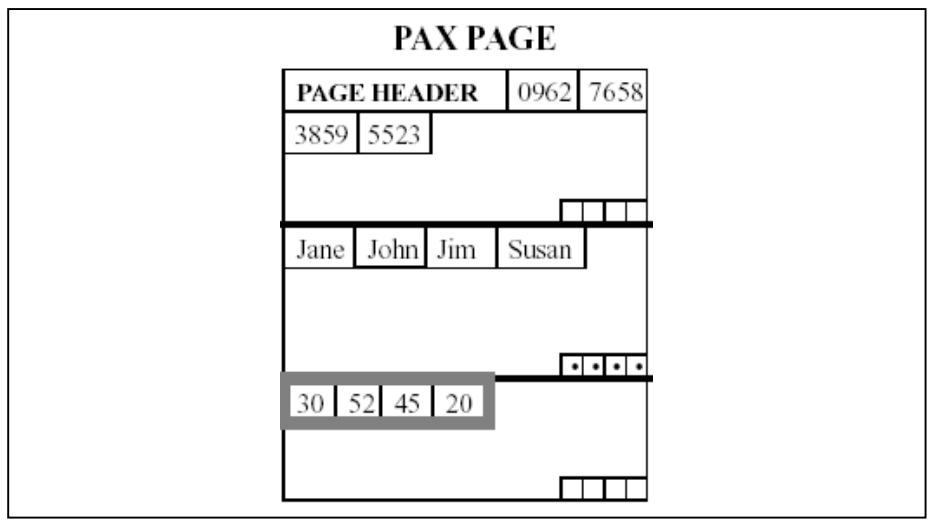

Na PAX, os dados são divididos em mini páginas, sendo que cada atributo é agrupado. A FIGURA 3 demonstra os agrupamentos para os atributos ID, Nome e Idade. Este tipo de divisão elimina referências desnecessárias à memória.

Esta estratégia pode ser incorporada em trabalhos futuros na atualização dos bancos de dados *open source* para melhoria de desempenho.

Citamos outras características de armazenamento de dados.

- O SGBD Oracle, por exemplo, armazena as tabelas em espaços conhecidos como "*tablespaces*" de acordo com o comportamento de cada tabela. Esta característica é muito útil para reduzir disputa entre os objetos do banco nos arquivos de dados.

- A criação e o dimensionamento dos segmentos de *"Rollback"*. Estes segmentos são utilizados para armazenar as informações que estão sendo atualizadas. Em um ambiente de *data warehousing*, no qual as atividades de carregamento exigem grandes transações, os segmentos de *rollback* devem ser grandes.

4.2.3 ÍNDICES

Os índices são responsáveis por significativos ganhos de desempenho na recuperação de informações em qualquer projeto de banco de dados e não somente em *data warehouse*.

Com a utilização de índices o SGBD percorre a estrutura de índice, obtém o endereço da linha e apresenta os dados. Sem um índice esta operação executaria uma varredura completa na tabela demandando maior processamento (CPU e E/S), conseqüentemente aumentando o tempo de resposta das consultas.

As pesquisas de índice são o segredo da otimização do tempo de resposta na maior parte das consultas e são usados sistematicamente em um *data warehouse* para melhorar seu desempenho de saída, segundo COREY et al, (2001, p. 399).

Diversos tipos de índices podem ser implementados, tais como: índices de única coluna, concatenados, exclusivos, baseados em função e índices de agrupamento.

Os índices de agrupamento, também chamados de *bitmap*, são muito utilizados em *data warehouse*. Alguns SGBDs, como o Sybase IQ, possuem implementações específicas deste tipo de índice dependendo da cardinalidade.

JEUNOT (2000) descreve que os índices de *bitmap* destinam-se a colunas de baixa cardinalidade, que contém um número limitado de valores.

Os SGBDs Oracle, IBM DB2, Sybase e Informix, por exemplo, implementam os índices de *bitmap* para uma cardinalidade de até 0,1% das linhas existentes na tabelas. Se tivermos numa tabela de 100.000 linhas o atributo de "UF" (unidade da federação) com 27 valores distintos, então, teremos uma cardinalidade de 27/100000 ou 0,027%, ou seja, para este atributo pode ser construído um índice de *bitmap*.

O SGBD Sybase IQ implementa vários tipos de índices de acordo com a cardinalidade, com valor baseado em *bitmap* ou não, além de índices específicos para os tipos das colunas utilizadas, como: DATE, TIMESTAMP e TIME.

O SGBD Informix XPS possui, além de índices de *bitmap*, índices baseados em seleção que são gerados sobre consultas previamente formatadas. Algumas aplicações de *data warehouse* utilizam este tipo de consulta como, por exemplo, desempenho diário de vendas de um determinado produto ou quantidade de ligações diárias para um determinado telefone.

WU e BUCHMANN (1998) definem que a idéia básica por trás de indexação de *bitmap* simples é utilizar um *string* de bits (0 ou 1) para indicar quando um atributo é igual a um valor especifico ou não. A FIGURA 5 mostra como um *bitmap* simples é implementado.

FIGURA 5 – EXEMPLO DE BITMAP SIMPLES

| TABELA T | | |
|---|---|---|
| ... | ATRIBA | ... |
| | a | |
| | b | |
| | c | |
| | c | |
| | a | |
| | ... | |

| Bitmap simples | | |
|---|---|---|
| Ba | Bb | Bc |
| 1 | 0 | 0 |
| 0 | 1 | 0 |
| 0 | 0 | 1 |
| 0 | 0 | 1 |
| 1 | 0 | 0 |
| | ... | |

No exemplo acima, a criação de um *bitmap* simples para o atributo ATRIBA com domínio {a, b, c} resulta em 3 vetores de *bits*, chamados Ba, Bb e Bc. Para Ba o bit é atualizado para 1 se o valor de ATRIBA=a, para Bb o bit é atualizado para 1 se o valor de ATRIBA=b e assim para cada valor do domínio do atributo ATRIBA.

Pode ser observado, por este exemplo que para cada valor do domínio é criado um vetor, que terá que ser gerenciado toda vez que o banco de dados cresce, além de operações neste índice, quando novos valores forem incorporados ao domínio. O problema para esta abordagem ocorre quando, por exemplo, uma tabela PRODUTOS contendo 12.000 diferentes produtos. Para esta tabela serão criados 12.000 vetores de *bits*.

WU e BUCHMANN (1998) propõem um índice de *bitmap* chamado *encoded bitmap* que melhora o desempenho em casos de atributos com grande domínio. Este índice utiliza uma cadeia de 0 e 1, ao invés de vetores para cada domínio do atributo, além de uma tabela de mapeamento e um conjunto de funções de recuperação.

No exemplo da FIGURA 5 o *encoded bitmap* em comparação com o *bitmap* simples seria implementado da seguinte forma:

- *bitmap* simples com 3 valores no domínio teria, portanto, 3 vetores de *bitmap*. Esta característica torna esta implementação mais rápida numa seleção única;

- *encoded bitmap* teria uma cadeia de 0 e 1 calculada através de, log2n, onde n é a quantidade de valores distintos do domínio, para o caso da FIGURA 7, $\log_2 3 = 2$ ;

A FIGURA 6 demonstra como ficaria a implementação do *encoded bitmap*.

FIGURA 6 – EXEMPLO DE *ENCODED BITMAP*

| TABELA T | | | | Encoded bitmap | | Tabela de mapeamento | |
|---|---|---|---|---|---|---|---|
| ... | ATRIBA | ... | | B1 | B0 | a | 00 |
| | a | | | 0 | 0 | b | 01 |
| | b | | | 0 | 1 | c | 10 |
| | c | | | 1 | 0 | | |
| | c | | | 1 | 0 | | |
| | a | | | 0 | 0 | | |
| | ... | | | ... | | | |

Se utilizarmos o exemplo da tabela de produtos, e utilizarmos o *encoded bitmap*, teríamos $\log_2 12000 = 14$, ou seja, ao invés de 12.000 vetores que seriam implementados pelo *bitmap* simples, teríamos um vetor de 12.000 elementos de 14 bits, mais a tabela de mapeamentos.

O SGBD IBM DB2 implementa este índice chamado pelo fabricante de *Encoded Vector Indexing* (EVI).

## 4.3 PODER DE PROCESSAMENTO

Um *data warehouse* possui como característica principal o apoio à tomada de decisão através dos seus altos volumes de dados carregados. Todo este volume de dados é carregado de tempos em tempos dependendo dos requisitos do negocio da empresa onde está sendo implementado. Os carregamentos podem chegar a vários milhões de registros por dia em alguns segmentos de mercado como o de telecomunicações e meteorologia, por exemplo.

Os equipamentos e aplicativos que suportam este tipo de carga de trabalho são extremamente caros haja vista este volume esperado.

Um dos fatores que demonstra o alto poder de processamento de um *data warehouse* além do carregamento é a complexidade das consultas. As consulta*s* que são submetidas ao banco de dados possuem um grau de complexidade muito superior ao de aplicações OLTP.

Nestas consultas podem ser observadas múltiplas junções, subconsultas e varreduras completas de tabelas, características estas pouco comuns ou nunca utilizadas em sistemas transacionais que possuem apenas consultas simples e nunca realizando operações de varredura completa de tabela.

Pelo grau de complexidade imposta por estas consultas de negócio, cálculos massivos são realizados exigindo muito poder de processamento, alem de muitas conexões simultâneas demandando todo tipo de informação.

Numa empresa, departamentos como marketing, vendas, financeiro, alta gerência e até alguns sistemas transacionais constantemente demandam informações do *data warehouse*, seja através de interface OLAP, conexão direta ou acesso através da SQL, por usuários mais experientes.

Os SGBDs que possuem versões para *data warehouse* implementam uma série de características que melhoram o desempenho da recuperação de informações. Estas características são operadores como o CUBE e ROLLUP que demonstram as informações em formato multidimensional, características estas anteriormente só implementadas pelas ferramentas de OLAP.

Outro fator que demonstra o poder de processamento de um ambiente de *data warehouse* é a quantidade de memória empregada. Considerando que algumas centenas de conexões são realizadas diariamente no banco de dados, a quantidade de memória deve ser grande e sua correta otimização dentro dos parâmetros do SGBD deve ser criteriosamente estudada, como descrito na seção de otimização.

Por fim, as operações de entrada e saída (E/S) exigem o máximo desempenho dos discos rígidos, pois a quantidade de dados envolvidos num estudo de negócios pode varrer os dados de vários dias, meses ou até anos.

Considerando o exemplo citado dos segmentos de telecomunicações e meteorológicos nos quais vários milhões de registros são gerados diariamente, o armazenamento de todo este volume de informações pode ocupar vários *terabytes* e a busca de informações, muito custosa, depende de estruturas que suportem operações massivas de entrada e saída de dados.

Numa comparação com um sistema transacional observa-se que em *data warehouse* as transações são longas, pois elas se caracterizam pelos carregamentos periódicos, portanto demandam operações massivas de entrada e saída. Num sistema transacional as transações são pequenas, pois sua finalidade é registrar as operações de um cliente, por exemplo.

Num *data* warehouse as consultas envolvem histórico demandando alto poder de entrada e saída de dados, memória e processamento. Num sistema transacional as consultas são simples buscando apenas a informação de um cliente para alguma informação pontual. Consultas complexas em sistemas transacionais podem comprometer o sistema como um todo e os DBAs que monitoram estes bancos de dados muitas vezes proíbem este tipo de operação.

Considerando a implementação de índices num *data warehouse,* a utilização de vários índices melhora o desempenho de consultas, mas aumenta o tempo de carregamento. O aumento deste tempo de carregamento torna-se aceitável devido ao ganho de desempenho em consultas. Em um sistema OLTP a utilização de muitos índices diminui o desempenho nas operações de DML (insert, update e delete) e a implementação de muitos índices num sistema deste tipo deve ser muito cuidadosa.

# 5 POSTGRESQL

O SGBD PostgreSQL é um aplicativo de código aberto que deriva seu desenvolvimento do SGBD Ingres.

O SGBD Ingres foi desenvolvido em 1977 na *University of California at Berkeley.* Em 1986 teve início o desenvolvimento de um SGBD objeto-relacional sobre o Ingres chamado de Postgres, sendo posteriormente chamado de PostgreSQL. O PostgreSQL é largamente conhecido como o mais avançado SGBD de código aberto DRAKE e WORSLEY (2002).

O PostgreSQL, por ser um software de código aberto possui na qualidade e na independência tecnológica fatores importantes; fatores estes defendidos pela comunidade de código aberto.

No fator qualidade é possível observar a constante preocupação da comunidade de software livre em estar desenvolvendo, bem como corrigindo os defeitos em comprometimento com o restante da comunidade.

O envolvimento de grandes empresas seja no desenvolvimento, como na utilização de um software livre, demonstra a qualidade dos produtos desenvolvidos.

Recentemente a Fujitsu anunciou a própria distribuição de SGBD baseada no PostgreSQL, chamado de Powergres Plus.

No fator independência tecnológica a comunidade de código aberto busca tornar o PostgreSQL o mais abrangente possível para as mais diversas plataformas e sistemas operacionais.

Dentre os SGBDs de código aberto o PostgreSQL é considerado o mais robusto, pois desde as suas versões mais antigas possui suporte completo as especificações da SQL, o que demonstra maior maturidade em comparação com outros SGBDs de código aberto.

## 5.1 CARACTERÍSTICAS

As principais características do PostgreSQL são:

- SGBD Objeto-relacional;

- Integridade referencial;
- Grande diversidade de interfaces (ODBC, JDBC, PHP, entre outros);
- Suporte as especificações da SQL92 e SQL99;
- Suporte a linguagens procedurais, seja a linguagem nativa ou outras linguagens como Perl, Python ou TCL;
- Controle de concorrência, evitando bloqueios de leitura quando existe uma escrita no banco de dados;
- Controle de escrita em *log* antes do disco.

O processamento de uma consulta PostgreSQL ocorre da seguinte forma, de acordo com LANE (2000), e pode ser visualizada na FIGURA 7:

1 - A consulta é submetida ao *parser* que verifica as definições dos objetos no dicionário de dados;

2 - É realizado a reescrita das consultas;

3 - O *planner* constrói um plano de execução orientado pela consulta reescrita e pelas estatísticas coletadas pelo DBA;

4 - É executado o plano criado pelo *planner*.

FIGURA 7 – PROCESSAMENTO DE CONSULTA NO POSTGRESQL

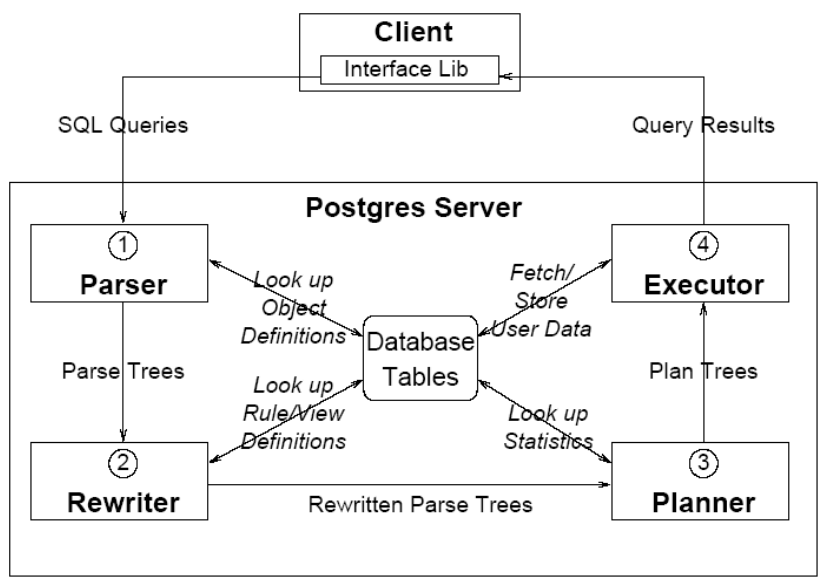

Até o início deste estudo em 2002 não haviam sido implementadas características específicas no PostgreSQL para suportar *data warehouse*.

Neste trabalho utilizamos a versão 7.4.2 do PostgreSQL. Na próxima versão, 8.x, serão implementadas duas características que aumentarão consideravelmente o desempenho do SGBD. As características são:

- Estatísticas de índice multi-coluna. Esta implementação poderá aumentar substancialmente o desempenho em muitas consultas que utilizamos neste trabalho, pois as grandes tabelas e as consultas complexas utilizam este tipo de índice;

- Espaços de tabela (*tablespaces)* que ajudam na organização de grandes bases de dados, distribuindo o armazenamento dos objetos. Esta distribuição do armazenamento poderá separar a localização dos dados e dos índices diminuindo a concorrência das operações de E/S.

Alguns estudos que apresentamos no capítulo 4 podem ser incorporados visando a melhoria do PostgreSQL considerando o desenvolvimento de um *data warehouse*, como:

- Implementação de estratégias de paginação PAX;

- Implementação de índices de *encoded bitmap*;

- Implementação do paralelismo *intra-query*, pois o PostgreSQL somente permite paralelismo *inter-query.*

Estudos e desenvolvimento sobre o PostgreSQL focando *data warehouse* podem trazer ao mercado uma boa opção, principalmente para os segmentos de mercado citados anteriormente que gostariam de implantar um *data warehouse* e não possuem recursos.

## 5.2 COMPARAÇÃO DE VALORES ENTRE SGBDs

O PostgreSQL, igualmente ao GNU/Linux, pode ser adquirido gratuitamente ou comercialmente se houver a necessidade de suporte específico.

Os valores gastos em SGBD para *data warehouse* são consideráveis, sendo que em alguns casos correspondem a quase 50% do valor total do sistema.

A comparação entre valores de SGBDs discriminados no TPC deve ser feita utilizando o TPC-H, pois os bancos de dados utilizados para *data warehouse* possuem características diferentes dos utilizados em outros tipos de aplicações.

É possível, portanto, demonstrar os valores despendidos na construção de *data warehouses* para a escala de 100 GB, utilizada neste estudo, apesar de existirem diferenças entre equipamentos. Também é possível observar o aumento no desempenho de cada ambiente através da métrica QphH na medida que aumenta o investimento. A métrica QphH ou *query-per-hour* é descrita no capítulo 6.

Os SGBDs Oracle e Teradata somente apresentam resultados a partir das escalas de 1000 GB e 3000 GB respectivamente.

O QUADRO 2 descreve os valores discriminados pelos fabricantes no TPC-H.

QUADRO 2 – COMPARAÇÃO ENTRE VALORES DE SGBD

| | Sybase IQ | MS SQL Server | IBM DB2 | Oracle | Teradata |
|---|---|---|---|---|---|
| **Sistema** | SunFire V240 Server | HP ProLiant ML370 G3 2P | IBM eServer xSeries 440 | HP 9000 Superdome Enterprise Server | NCR 5350 |
| **Custo total (US$)** | 45.021 | 38.683 | 436.680 | 5.249.167 | 16.937.451 |
| **Sistema Operacional** | Sun Solaris 9 | Microsoft Windows Server 2003 Standard Edition | Microsoft Windows Server 2003 Enterprise Edition | HP UX 11.i 64-bit | MP-RAS 3.02.00 |
| **Versão SGBD** | Sybase IQ 12.5 | Microsoft SQL Server 2000 Enterprise Edition | IBM DB2 UDB 8.1 | Oracle 9i Database Enterprise Edition v9.2.0.2.0 | Teradata V2R5.0 |
| **CPU** | Sun UltraSPARC[TM] IIIi Cu 1 GHz | Intel Xeon 3.06 GHz | Intel Xeon MP 2.0 GHz | HP PA-RISC 8700 750MHz | Intel Xeon 2.8GHz |
| **Cluster** | N | N | N | N | Y |
| **Custo total SGBD (US$)** | 19.920 | 17.189 | 251.322 | 806.000 | 2.205.829 |
| **# CPU** | 2 | 2 | 8 | 64 | 128 |
| **QphH** | 1.124,4 | 1.386,0 | 3.342,7 | 25.805,4 | 79.528,0 |
| **Escala** | 100 GB | 100 GB | 100 GB | 1000 GB | 3000 GB |

Fonte: TPC-H  Data acesso: 24/06/03

A escolha de uma plataforma que ofereça um desempenho melhor muitas vezes pode ser custoso, podendo ser observado em uma comparação entre as plataformas com SGBDs SQL Server e DB2 para a escala de 100 GB ilustrado pelo QUADRO 2. A plataforma com DB2 possui um desempenho 2,4 vezes maior que a plataforma com SQL Server, mas possui um custo 11,2 vezes maior.

Os valores do custo total de SGBDs descritos acima consideram, dependendo da plataforma, o seguinte:

- Sybase
  - o Sybase IQ-M Single App Srv – 2 Cpu's;
  - o Suporte de três anos Sybase;
  - o Desconto (licença e suporte);

- MS SQL Server
  - o MS SQL Server 2000 Enterprise Edition;
  - o Pacote de suporte do banco de dados;
  - o Licenças dos clientes;

- IBM DB2
  - o DB2 UDB ESE 8.1 Licença e manutenção 1 ano;
  - o DB2 UDB ESE 8.1 suporte 1 ano;
  - o Pacote de suporte do banco de dados;

- Oracle
  - o Oracle9i, Enterprise Edition Release 2 com usuários nomeados por três anos;
  - o Particionamento com usuários nomeados;
  - o Suporte Oracle Database por três anos;

- Teradata
  - o Teradata/DB V2R5.0 WM 5350 MP-RAS;
  - o Manutenção de três anos;

No resultado apresentado pelo Teradata, o SGBD e o sistema operacional são comercializados em conjunto.

# 6 METODOLOGIA

## 6.1 VISÃO GERAL

Nesta seção será descrita a metodologia desenvolvida pelo TPC, observando os itens pertinentes ao tipo de implementação utilizada neste trabalho. A especificação completa poderá ser encontrada no próprio *website* do TPC (www.tpc.org), pois esta seção destina-se apenas a descrever a metodologia de forma genérica.

Serão também apresentadas as diferenças entre o TPC-H e a metodologia OSDL DBT3 que foi utilizada neste trabalho.

Até o inicio desta dissertação a metodologia TPC-H estava na versão 2.0.0. A versão TPC-H 2.1.0, que segue a que utilizamos, somente define novas escalas de tamanho, o que não altera o resultado do que foi descrito.

TPC *Benchmark* H é compreendido de consultas de negócio projetadas para exercitar as funcionalidades de um sistema buscando representar aplicações complexas de análises de negócios. Estas consultas possuem um contexto realista demonstrando a atividade de um fornecedor de vendas. O escopo escolhido visa ajudar o leitor a relacionar intuitivamente os componentes do *benchmark* quando estiver analisando as consultas TPC (2003).

A metodologia não tenta modelar um segmento de mercado em específico, mas atividades que sejam comuns a vários segmentos. A metodologia também prevê requisitos quanto à transparência no acesso aos dados. Estes requisitos retiram da consulta qualquer necessidade de saber a localização dos dados.

Na FIGURA 8 está o esquema empregado nesta metodologia indicando a cardinalidade, o fator de escala (SF) e uma convenção para nomear os atributos de cada tabela.

FIGURA 8 – ESQUEMA TPC-H 2.0.0

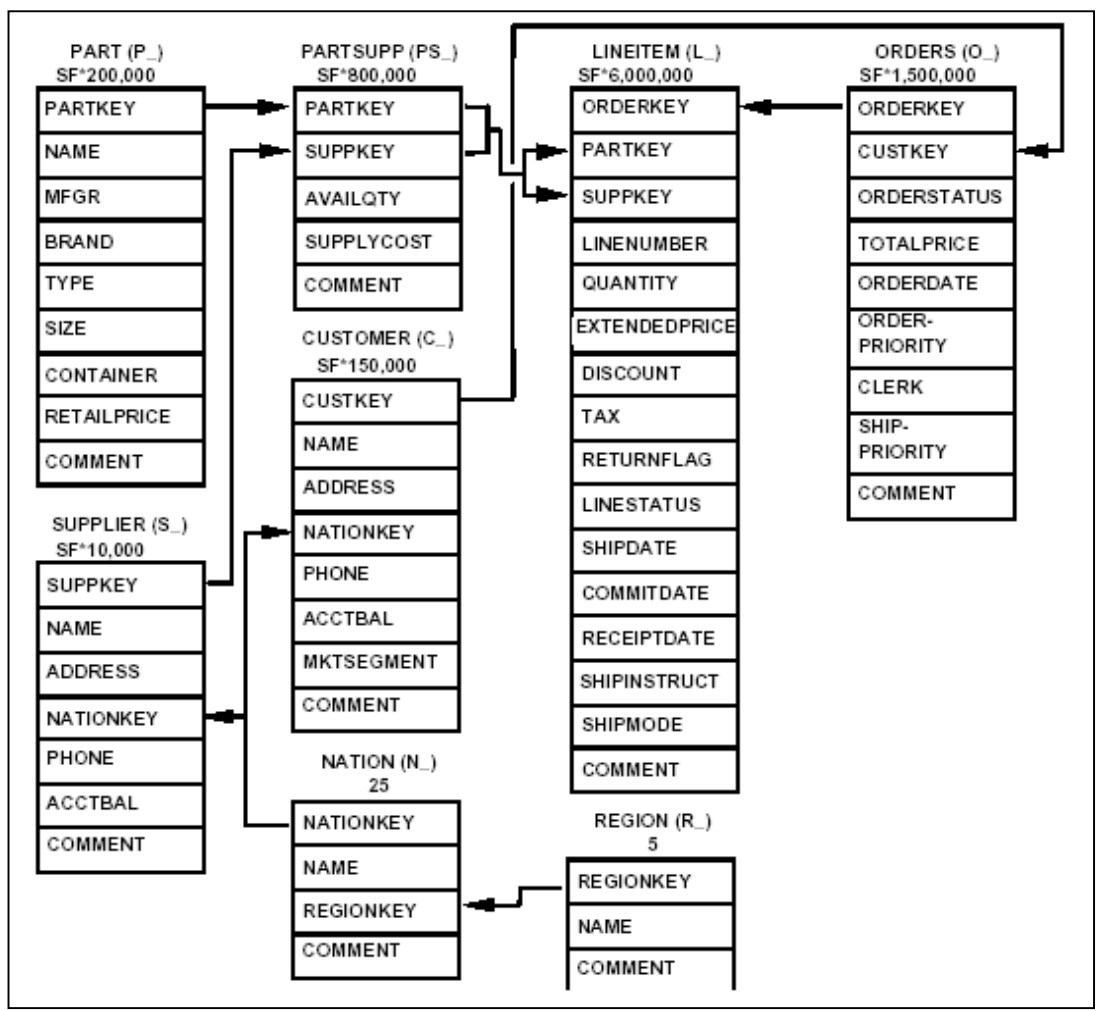

A metodologia emprega o esquema delineado na FIGURA 8, contendo o *layout* das tabelas, bem como a descrição de chave primária e de chave estrangeira, as restrições de cada atributo e as consultas que podem ser vistas no ANEXO 3.

O TPC também disponibiliza um gerador de dados para popular o banco de dados para o *benchmark*, sendo este configurável de acordo com a plataforma e a escala utilizada.

O gerador de dados chama-se DBGEN e foi escrito em ANSI C numa tentativa de ser portável para várias plataformas.

O DBGEN possui em linhas de comando opções para gerar os dados diretamente no banco de dados ou em arquivo texto. Na opção de geração de arquivo texto é possível gerar para cada tabela diversos arquivos, pois tabelas como a LINEITEM são muito grandes e a criação de um arquivo com tamanho muito grande pode não ser permitido pelo sistema operacional.

Na opção de geração de arquivos texto, o tempo de criação dos arquivos não é contado, somente o carregamento dos dados no banco de dados.

A outra opção de geração dos dados é realizada diretamente no banco de dados onde o tempo de geração e carregamento é contado, pois como o DBGEN, na medida que gera os dados também os carrega no banco de dados, o tempo é contado unicamente.

As escalas de tamanho do TPC-H, ilustrado na FIGURA 8 como SF, estão divididas em 1 GB, 10 GB, 30 GB, 100 GB, 300 GB, 1.000 GB, 3.000 GB e 10.000 GB. O TPC reconhece que é necessária a criação de escalas maiores e para a versão TPC-H 2.1.0 incorpora as escalas 30.000 GB e 100.000 GB.

Neste estudo utilizamos as escalas de 1GB e 100GB. Para a escala de 100GB, o tamanho físico do banco de dados e a cardinalidade das tabelas excluindo os índices, está no QUADRO 3.

QUADRO 3 – TAMANHO DO BANCO DE DADOS

| TABELAS | | | |
|---|---|---|---|
| Tabela | Tuplas | Páginas | Tam (bytes) |
| REGION | 5 | 1 | 8.192 |
| PART | 20.000.000 | 511.629 | 4.191.264.768 |
| SUPPLIER | 1.000.000 | 25.840 | 211.681.280 |
| PARTSUPP | 80.000.000 | 1.892.470 | 15.503.114.240 |
| CUSTOMER | 15.000.000 | 421.512 | 3.453.026.304 |
| ORDERS | 150.000.000 | 3.188.114 | 26.117.029.888 |
| LINEITEM | 600.038.000 | 15.797.338 | 129.411.792.896 |
| NATION | 25 | 1 | 8.192 |
| Total Tabelas | 866.038.030 | 21.836.905 | 178.887.925.760 |
| Total Índices | - | 7.721.629 | 63.255.584.768 |
| Total BD | 866.038.030 | 29.558.534 | 242.143.510.528 |

O QUADRO 4 ilustra o tamanho dos índices em páginas e *bytes.*

QUADRO 4 – TAMANHO DO BANCO DE DADOS (ÍNDICES)

| ÍNDICES | | |
|---|---|---|
| relname | relpages | Tam (bytes) |
| nation_pkey | 2 | 16.384 |
| part_pkey | 54.840 | 449.249.280 |
| supp_pkey | 2.745 | 22.487.040 |
| ps_pkey | 219.355 | 1.796.956.160 |
| customer_pkey | 41.131 | 336.945.152 |
| orders_pkey | 411.289 | 3.369.279.488 |
| i_l_orderkey | 1.645.244 | 13.477.838.848 |
| i_n_nationkey | 2 | 16.384 |
| i_o_orderdate | 411.289 | 3.369.279.488 |
| i_l_shipdate | 1.645.244 | 13.477.838.848 |
| i_l_receiptdate | 1.645.244 | 13.477.838.848 |
| i_l_partkey | 1.645.244 | 13.477.838.848 |

Como neste trabalho o banco de dados é gerado através de arquivos texto, o tempo de geração dos arquivos não é computado no tempo de carregamento.

TPC (2003) define que as principais regras de implementação são:

- O banco de dados deve ser implementado usando um SGBD comercialmente disponível;
- No final do teste de carregamento as tabelas devem ter o número exato de linhas para o fator de escala;
- Os nomes de tabelas e colunas devem estar em conformidade com os descritos pelo TPC;
- Não é permitida a inclusão de colunas além das especificadas;
- O banco de dados deve permitir a inclusão de dados arbitrários em conformidade com o tipo do dado e restrições opcionais definidas;
- Particionamento horizontal é permitido e sendo baseado num campo, este deve ser: chave primária, chave estrangeira ou única coluna de data;

- Alguns particionamentos requerem o uso de diretivas que especifiquem os valores dos campos de particionamento. Se estas diretivas forem utilizadas elas devem satisfazer as seguintes condições:
  - o Não podem confiar em nenhum conhecimento dos dados nas tabelas, exceto os valores mínimo e máximo do campo de particionamento;
  - o Dentro das limitações de divisão de números inteiros, devem ser definidas partes iguais dentro do conjunto entre os valores mínimo e máximo;
  - o As diretivas devem permitir a inclusão de valores na coluna de particionamento que estejam fora do conjunto entre os valores mínimo e máximo;

Para o carregamento do banco de dados foi criado um *shell script* que conta, inclusive, o tempo deste carregamento.

No tempo de carregamento também são contados os tempos de criação de índices, validações de restrições, bem como coleta de estatísticas para a otimização do sistema.

Para verificar se o carregamento foi completado com sucesso deve ser feita uma contagem dos registros carregados em cada tabela. Esta contagem deve ser comparada com o cálculo demonstrado na FIGURA 8 onde SF multiplicado pela cardinalidade resulta na quantidade de registros a serem carregados em cada tabela.

Como parte do *benchmark,* o TPC seleciona 22 consultas de recuperação de dados e 2 consultas de atualização de dados. As consultas estão descritas na documentação juntamente com o contexto de negócio, o código SQL, os parâmetros de substituição, *c*onsulta de validação e o resultado desta validação.

O TPC (2003) descreve que as consultas selecionadas possuem as seguintes características:
- Alto grau de complexidade;
- Vários tipos de acesso;
- Possuem uma natureza *ad-hoc*;

- Examinam uma grande porcentagem dos dados disponíveis;
- Diferem uma das outras;
- Possuem parâmetros que se alteram entre as execuções.

Estas consultas provêem os seguintes resultados:
- Preço e promoções;
- Suprimentos e gerenciamento de demanda;
- Lucro e gerenciamento de rendimentos;
- Estudo de satisfação do cliente;
- Estudo de participação de mercado;
- Gerenciamento de embarque de mercadorias.

O teste de desempenho consiste em duas execuções que são o teste de poder (*power test*) e teste de produtividade *(throughput test)*.

Cada execução é composta por: consulta, conjunto de consultas, seqüência de consultas, seqüência de atualizações e as sessões, onde:
- Consulta é uma das vinte e duas consultas TPC-H descritas;
- Seqüência de consultas é a execução seqüencial de um conjunto de 22 consultas;
- Seqüência de atualizações são consultas de atualização submetidas seqüencialmente por um programa em lote;
- Sessão é um processo capaz de executar o conjunto de consultas ou de atualizações.

A comunicação entre o sistema sob teste e o programa que irá submeter às consultas deverá considerar uma sessão para cada seqüência de consultas ou atualizações.

O TPC (2003) especifica que de acordo com a carga de trabalho a configuração do sistema sob teste deverá ser baseado nas documentações disponíveis. Esta carga de trabalho é caracterizada por:
- Varredura seqüencial de grandes volumes de dados;
- Agregações de grandes volumes de dados;

- Junções de múltiplas tabelas;
- Ordenações.

Algumas características e parâmetros de configuração que alteram o desempenho do sistema e que os administradores do banco de dados devem saber e decidir quando usar são:
- Conhecimento da carga de trabalho;
- Conhecimento do modelo lógico e físico do banco de dados;
- Acesso à documentação do banco de dados e do sistema operacional;
- Nenhum conhecimento do funcionamento interno dos produtos utilizados além dos providos pela documentação externa.

## 6.2 REGRAS DE EXECUÇÃO

Como regras gerais de execução, o TPC (2003) descreve:
- Cada seqüência de consultas deve corresponder a uma sessão;
- Paralelismo *intra-query* não está restringido;
- O teste de desempenho do sistema será dividido em:
  - o Teste de poder no qual é medido tempo da execução simples de cada consulta como usuário único. Neste teste as consultas de atualização também são testadas separadamente antes e depois da execução das consultas de seleção;
  - o Teste de produtividade que demonstra a habilidade do sistema de processar as consultas no menor tempo. Neste teste, dependendo da escala, devem ser executadas múltiplas seqüências de consultas ou *streams* de consultas, como demonstrado no QUADRO 5.

QUADRO 5 – SEQUÊNCIA DE CONSULTAS POR ESCALA

| SF | S(Streams) |
|---|---|
| 1 | 2 |
| 10 | 3 |
| 30 | 4 |
| 100 | 5 |
| 300 | 6 |
| 1000 | 7 |
| 3000 | 8 |
| 10000 | 9 |

No QUADRO 5 o fator de escala corresponde à legenda SF e a quantidade de seqüências de consultas que deverão ser submetidas ao SGBD correspondem à legenda S(Streams). A execução de cada consulta na seqüência de consultas deve ser realizada uma após a outra.

Neste teste, a execução das consultas de seleção deverá ocorrer em paralelo com uma consulta de atualização.

O TPC impõe que o sistema sob teste deverá ser o mesmo para os dois testes, ou seja, após o teste de poder não deverá existir qualquer modificação para a execução do teste de *throughput*.

As regras de submissão das consultas de seleção e atualização, ou seja, intervalo entre cada consulta, seqüência e medida do tempo do intervalo estão definidas na especificação.

A especificação também define como deverá ser feita a implementação do gerador e do programa de execução do *benchmark*, mas como estaremos utilizando o DBGEN e o QGEN como gerador de dados e programas de execução das consultas respectivamente, então não estamos relatando esta parte da especificação.

## 6.3 MÉTRICAS

As métricas definidas para o *benchmark* são:

- TPC-H *Power;*
- TPC-H *Query-por-hora* ou produtividade*;*
- TPC-H *Price/Performance* ou Preço / Desempenho;
- Disponibilidade do sistema.

As três últimas métricas são consideradas as principais.

O TPC-H *Power* é uma métrica utilizada na formula do TPC-H *Query-por-hora* e está definida no QUADRO 6.

QUADRO 6 – MÉTRICA TPC-H *POWER@SIZE*

$$\text{TPC-H Power@Size} = \frac{3600 * SF}{\sqrt[i=22]{\prod_{i=1}^{i=22} QI(i,0) * \prod_{j=1}^{j=2} RI(j,0)}}$$

Onde:

- SF é o fator de escala;
- QI é o intervalo em segundos de cada consulta de seleção;
- RI é o intervalo em segundos de cada consulta de atualização.

O *Throughput@Size* é uma métrica utilizada na formula do TPC-H *Query-por-hora* e está definida no QUADRO 7.

QUADRO 7 – MÉTRICA TPC-H *THROUGHPUT@SIZE*

$$\text{TPC-H Throughput@Size} = (S * 22 * 3600) / T_S * SF$$

Onde:

- S é a quantidade da seqüência de consultas que depende do fator de escala;

- T é a medição do intervalo de tempo;
- SF é o fator de escala.

O TPC-H *Query-por-hora* é definido no QUADRO 8 pela fórmula:

QUADRO 8 – MÉTRICA TPC-H *QUERY-POR-HORA*

$$QphH@Size = \sqrt{Power@Size * Throughput@Size}$$

Onde:

- *Power@Size* é definido pela métrica TPC-H *Power;*
- *Throughput@Size* é definido pela métrica de acordo com o fator de escala.

O TPC-H Preço / Desempenho é definido no QUADRO 9 pela fórmula:

QUADRO 9 – MÉTRICA TPC-H PREÇO / DESEMPENHO

$$TPC\text{-}H\ Price\text{-}per\text{-}QphH@Size = \$/QphH@Size$$

Onde:

- $ é o valor de todo o sistema;
- QphH@size é definido pela medida TPC-H *Query-por-hora.*

6.4 DIFERENÇAS ENTRE TPC-H E OSDL DBT3

O OSDL DBT3 é uma adaptação do TPC-H para testar o sistema operacional GNU/Linux.

A decisão de utilizar o DBT3 reside no fato do PostgreSQL não resolver de maneira eficaz algumas consultas como ficou demonstrado nos resultados.

Citamos abaixo os fatores que contribuíram para a utilização do DBT3:

- Possibilidade de reescrita das consultas, pois o PostgreSQL não resolve as consultas de maneira eficaz como ficou demonstrado. A reescrita é proibida pelo TPC;

- Pode ser utilizado qualquer fator de escala e não somente os utilizados pelo TPC-H. Apesar disto foram utilizados neste trabalho as escalas 1GB e 100GB.

- No TPCH algumas consultas são restringidas no retorno de linhas e esta restrição não é aplicada pelo DBT3. Neste trabalho utilizamos a mesma opção do DBT3;

- Não é realizado o teste de ACID no DBT3;

- Não é utilizada a métrica de preço/desempenho;

- O DBT3 possui a métrica Composite cujo cálculo é o mesmo ilustrado no QUADRO 8 para a métrica TPC-H *Query-per-hour*.

# 7 CONFIGURAÇÕES E IMPLEMENTAÇÕES

Nesta seção estão descritas as configurações realizadas no DBGEN e QGEN.

## 7.1 *SCRIPTS* E CONFIGURAÇÕES DO DBGEN E QGEN

Na geração dos dados através do DBGEN a configuração e documentação disponibilizada não possuíam parâmetros específicos para a utilização do GNU/Linux, sendo os sistemas operacionais destino os seguintes: ATT, DIGITAL, DOS, HP, IBM, MVS, ICL, SGI, SUN, TANDEM, U2200 e VMS. Para podermos utilizar o GNU/Linux fizemos uma alteração baseada nos experimentos e verificação dos arquivos gerados. Esta alteração foi feita no arquivo "config.h" e está descrita na FIGURA 9.

FIGURA 9 – CONFIGURAÇÃO DBGEN ARQUIVO CONFIG.H

```
#ifdef LINUX
#define EOL_HANDLING
#define STDLIB_HAS_GETOPT
#endif /* LINUX */
```

Onde EOL_HANDLING não considera separador de final de linha e STDLIB_HAS_GETOPT previne conflitos da função *getopt()*.

Com a utilização de EOL_HANDLING não foi necessária a criação de mais uma coluna em cada tabela, pois o finalizador de cada linha nos arquivos texto era gerado com o mesmo caractere de delimitador de colunas, e com a utilização desta função o final de linha fica delimitado com caractere específico para este fim. Foi necessária a utilização desta função, pois a metodologia proíbe a modificação da estrutura das tabelas.

Após a alteração e compilação do DBGEN, foi gerado o banco de dados para a escala de 1GB e 100GB nos quais realizamos os testes que tem seus resultados descritos em seção própria.

A geração dos dados foi realizada através de um *script*. Este *script* está detalhado na FIGURA 10.

FIGURA 10 – COMANDOS DE GERAÇÃO DOS DADOS

```
#!/bin/ksh
./dbgen -s 100 -f -T P -v
./dbgen -s 100 -f -T s -v
./dbgen -s 100 -f -T r -v
./dbgen -s 100 -f -T n -v
./dbgen -s 100 -f -T c -v
# PARTSUPP
./dbgen -s 100 -S 1 -C 4 -f -T S -v
./dbgen -s 100 -S 2 -C 4 -f -T S -v
./dbgen -s 100 -S 3 -C 4 -f -T S -v
./dbgen -s 100 -S 4 -C 4 -f -T S -v
# ORDERS
./dbgen -s 100 -S 1 -C 4 -f -T O -v
./dbgen -s 100 -S 2 -C 4 -f -T O -v
./dbgen -s 100 -S 3 -C 4 -f -T O -v
./dbgen -s 100 -S 4 -C 4 -f -T O -v
# LINEITEM
integer i
((i = 1))
while((i < 31))
do
  ./dbgen -s 100 -S $i -C 30 -f -T L -v
  ((i+=1))
done
exit
```

As tabelas PARTSUPP, ORDERS e LINEITEM foram geradas e particionadas por causa do tamanho. O particionamento buscava gerar arquivos menores de 2GB cada, pois arquivos maiores que 2GB não são permitidos pelo kernel versão 2.4.x.

O carregamento dos dados é realizado pelo conjunto de comandos descritos na FIGURA 11.

FIGURA 11 – COMANDOS DE CARREGAMENTO DOS DADOS

```ksh
#!/bin/ksh
echo "----------" > log.txt
echo "Inicio " >> log.txt
echo |date +%H:%M:%S >> log.txt
integer j

echo "LINEITEM" >> log.txt
echo |date +%H:%M:%S >> log.txt

((j = 1))
while((j<31))
do
     cat lineitem.tbl.$j | psql -c "copy lineitem from stdin delimiters '|'" -d
db_eduardo
        rm lineitem.tbl.$j
        echo "lineitem" $j >> log.txt
     ((j+=1))
done

echo "ORDERS" >> log.txt
echo |date +%H:%M:%S >> log.txt
((j = 1))
while((j<5))
do
     cat orders.tbl.$j | psql -c "copy orders from stdin delimiters '|'" -d db_eduardo
        rm  orders.tbl.$j
     ((j+=1))
done

echo "REGION " >> log.txt
echo |date +%H:%M:%S >> log.txt
cat region.tbl | psql -c "copy region from stdin delimiters '|'" -d db_eduardo
rm region.tbl

echo "NATION "  >> log.txt
echo |date +%H:%M:%S >> log.txt
cat nation.tbl | psql -c "copy nation from stdin delimiters '|'" -d db_eduardo
rm nation.tbl

echo "PART "  >> log.txt
echo |date +%H:%M:%S >> log.txt
cat part.tbl | psql -c "copy part from stdin delimiters '|'" -d db_eduardo
rm part.tbl
```

```
echo "SUPPLIER " >> log.txt
echo |date +%H:%M:%S >> log.txt
cat supplier.tbl | psql -c "copy supplier from stdin delimiters '|'" -d db_eduardo
rm supplier.tbl

echo "PARTSUPP " >> log.txt
echo |date +%H:%M:%S >> log.txt
((j = 1))
while((j < 5))
do
 cat partsupp.tbl.$j | psql -c "copy partsupp from stdin delimiters '|'" -d db_eduardo
      ((j+=1))
done
rm partsupp*

echo "CUSTOMER" >> log.txt
echo |date +%H:%M:%S >> log.txt
cat customer.tbl | psql -c "copy customer from stdin delimiters '|'" -d db_eduardo
rm customer.tbl

echo "Final " >> log.txt
echo |date +%H:%M:%S >> log.txt
exit
```

Os comandos de carregamento enviam para um arquivo de *log* os tempos de carregamento de cada tabela e carregam os arquivos através do aplicativo psql do PostgreSQL

Este conjunto de comandos realiza o carregamento dos dados no banco de dados, detalhando os tempos deste carregamento para cada tabela. Os tempos de carregamento também devem considerar a criação dos índices e restrições do banco de dados.

Para criar as consultas descritas pelo TPC foi utilizado o aplicativo QGEN também disponibilizado pelo TPC.

O QGEN foi desenvolvido em ANSI C buscando ser portável para várias plataformas diferentes. Em sua estrutura existem as consultas modelo que são utilizadas para a geração das consultas definitivas. O QGEN gera as consultas de acordo com as seqüências estipuladas pela metodologia, substituindo os valores nas consultas modelo e gerando as seqüências de consultas.

As variáveis de ambiente configuradas para a correta execução do DBGEN e QGEN no GNU/Linux estão descritas na FIGURA 12.

FIGURA 12 – VARIÁVEIS DE AMBIENTE

```
DSS_CONFIG=/home/eduardo/tpch/appendix/dbgen
DSS_DIST=dists.dss
DSS_PATH=/home/eduardo/tpch/appendix/files
DSS_QUERY=/home/eduardo/tpch/appendix/queries
```

As variáveis descrevem o seguinte:

DSS_CONFIG é o diretório onde é encontrado o arquivo que contém os parâmetros de configuração;

DSS_DIST é o arquivo contendo os valores que serão substituídos nos textos das consultas modelo;

DSS_PATH é o diretório onde serão criados os arquivos texto gerados pelo DBGEN, e

DSS_QUERY é o diretório onde estão as consultas modelo para a geração das consultas definitivas.

O script para geração das consultas utilizando o QGEN está descrito na FIGURA 13.

FIGURA 13 – COMANDOS DE GERAÇÃO DAS CONSULTAS

```
./qgen -l ../files/stream00.par -p 0 -s 100 -c -i $DSS_QUERY/init.sql -t
$DSS_QUERY/complete.sql > ../files/stream00.sql
./qgen -l ../files/stream01.par -p 1 -s 100 -c -i $DSS_QUERY/init.sql -t
$DSS_QUERY/complete.sql > ../files/stream01.sql
./qgen -l ../files/stream02.par -p 2 -s 100 -c -i $DSS_QUERY/init.sql -t
$DSS_QUERY/complete.sql > ../files/stream02.sql
./qgen -l ../files/stream03.par -p 3 -s 100 -c -i $DSS_QUERY/init.sql -t
$DSS_QUERY/complete.sql > ../files/stream03.sql
./qgen -l ../files/stream04.par -p 4 -s 100 -c -i $DSS_QUERY/init.sql -t
$DSS_QUERY/complete.sql > ../files/stream04.sql
./qgen -l ../files/stream05.par -p 5 -s 100 -c -i $DSS_QUERY/init.sql -t
$DSS_QUERY/complete.sql > ../files/stream05.sql
```

Os comandos do QGEN ilustrados na FIGURA 13 indicam o arquivo com os parâmetros utilizados em cada consulta (opção -l), a quantidade de *streams* (opção -p), a escala utilizada (opção -s), a inclusão de um cabeçalho (opção -i) e rodapé (opção -t) no texto de cada consulta como ilustrado no ANEXO 3.

As configurações utilizadas nos arquivos de configuração do TPC para incluir os comandos do PostgreSQL nos arquivos "tpcd.h" e "makefile" foram:

FIGURA 14 – CONFIGURAÇÃO POSTGRESQL ARQUIVO TPCD.H

```
#ifdef POSTGRESQL
#define GEN_QUERY_PLAN  "EXPLAIN"
#define START_TRAN    "START TRANSACTION;"
#define END_TRAN      "COMMIT;"
#define SET_OUTPUT     ""
#define SET_ROWCOUNT    ""
#define SET_DBASE     "psql -d %s"
#endif
```

FIGURA 15 – CONFIGURAÇÃO PARA O ARQUIVO MAKEFILE

```
DATABASE = POSTGRESQL
MACHINE = LINUX
WORKLOAD = TPCH
```

Na figura 15, a variável DATABASE mapeia a configuração do SGBD no arquivo "tpcd.h" e a variável MACHINE mapeia a configuração do sistema operacional no arquivo "config.h".

O arquivo "makefile" compila e gera os executáveis utilizando as configurações descritas nos arquivos "tpcd.h" e "config.h"

A implementação das funções de atualização (RF1 e RF2) foi escrita em código JAVA e os núcleos destes códigos estão descritos na seção ANEXOS. A função "RF1" realiza diversas inclusões no banco de dados através de uma formula descrita pelo TPC-H. A função RF2 realiza exclusões no banco de dados. As duas funções, além das seqüências de consultas, buscam manter o banco de dados em condições reais para que o *benchmark* possa apresentar resultados confiáveis.

A função RF1 realiza, de acordo com descrição do TPC, entre uma e sete inclusões na tabela LINEITEM para cada inclusão na tabela ORDERS. Estas inclusões são realizadas através dos dados nos arquivos gerados pelo DBGEN para as funções de atualização.

A função RF2 realiza exclusões através dos dados dos arquivos gerados pelo DBGEN. A exclusão é realizada buscando as chaves disponibilizadas nestes arquivos.

# 8 RESULTADOS

A execução dos testes de desempenho foi iniciada a partir do TPC-H, que é a metodologia base deste estudo. Após o *benchmark* TPC-H executamos o DBT3, por motivos descritos na seção de metodologia, aferidos através dos resultados exibidos nesta seção.

Inicialmente foi utilizada a escala de 100 GB e na medida que os resultados foram gerados executamos o TPC-H de 1 GB e o DBT3 de 1 GB.

O ambiente preparado para o *benchmark* utilizou a seguinte configuração:

- OSDL DBT3 versão 1.4 e TPC-H versão 2.0.0;
- Fator de escala de 1GB e 100GB;
- Carregamento realizado utilizando arquivos (*flat file*).

Software
- PostgreSQL 7.4.2
- Mandrake GNU/Linux 64bits (kernel 2.6.5)
- Java SDK 1.4.2_04
- PostgreSQL JDBC pg74.1jdbc3.jar
- Utilitários TPC-H (DBGEN and QGEN)

Hardware
- Dual Opteron 64bits Model 240 1.4GHz
- 4 GB RAM
- 960 GB Disk em RAID 0

O gasto com hardware adquirido em março de 2.004 foi R$ 16.159,00.

O gasto com software seria a partir de US$ 100, mas como se tratava de um estudo foi realizado um *download* do sistema operacional não representando custos. O sistema operacional é o único software que gera custo de aquisição caso este ambiente seja utilizado, pois se trata de uma nova tecnologia. Até o inicio deste estudo apenas distribuições pagas de GNU/Linux para Opteron 64bits foram encontradas. As distribuições são RedHat, Suse e Mandrake.

Como descrito anteriormente, os programas para carregamento do banco de dados e execução do *benchmark* foram escritos para o TPC-H em Java e Shell script e para o DBT3 utilizamos o pacote disponibilizado pela OSDL em seu *website*.

Foi utilizado JavaSDK versão 32-bits, pois a Sun Microsystems não possuía, até o inicio dos testes, versão estável para Opteron 64-bits, e por este motivo tivemos que compilar o kernel 2.6.5 com suporte a 32-bits.

O carregamento do banco de dados foi realizado três vezes utilizando kernel versões 2.4.24 e 2.6.5 para medir as diferenças e para poder extrair o máximo de desempenho do equipamento e sistema operacional. Primeiramente criamos um *script* monolítico, ou seja, o carregamento não foi distribuído para os dois processadores.

A partir dos tempos de carregamento tomamos a decisão de utilizar o kernel versão 2.6.5 e dividimos o *script* de modo a utilizar os dois processadores.

A divisão realizada carrega, constrói e otimiza a tabela LINEITEM num *script* separado e o resto do banco de dados em outro *script*. A tabela LINEITEM é a maior tabela do sistema como demonstrado no QUADRO 3 na seção de metodologia.

A divisão do *script* resultou num ganho significativo no carregamento, os tempos de carregamento são descritos no QUADRO 10.

QUADRO 10 – TEMPOS DE CARREGAMENTO

| | | | | Tempo(Hr) | | |
|---|---|---|---|---|---|---|
| Distribuição | Kernel | Scala(GB) | Script | Data load | PK and index | Total load time |
| Debian 32bits | 2.4.24 | 100 | Monolitico | 11:37:59 | 12:22:04 | 24:00:04 |
| Mandrake 64bits | 2.6.5 | 100 | Monolitico | 07:24:23 | 13:42:31 | 21:06:54 |
| Mandrake 64bits | 2.6.5 | 100 | Distribuído | 06:14:30 | 09:58:31 | 16:13:01 |
| Mandrake 64bits | 2.6.5 | 1 | Distribuído | 00:03:40 | 00:03:39 | 00:07:19 |

Com a divisão do *script* e a escolha de uma distribuição GNU/Linux de 64bits em kernel versão 2.6.x o tempo de carregamento foi 32,43% menor em comparação com o ambiente em 32bits. Comparando com o mesmo ambiente, mas utilizando *scrpit* monolítico o tempo de carregamento ficou 23,19% menor.

Após o carregamento foram iniciados os testes.

O resultado do power test do benchmark TPC-H para a escala de 100GB está demonstrado na FIGURA 16.

## FIGURA 16 – RESULTADO *POWER TEST* TPC-H

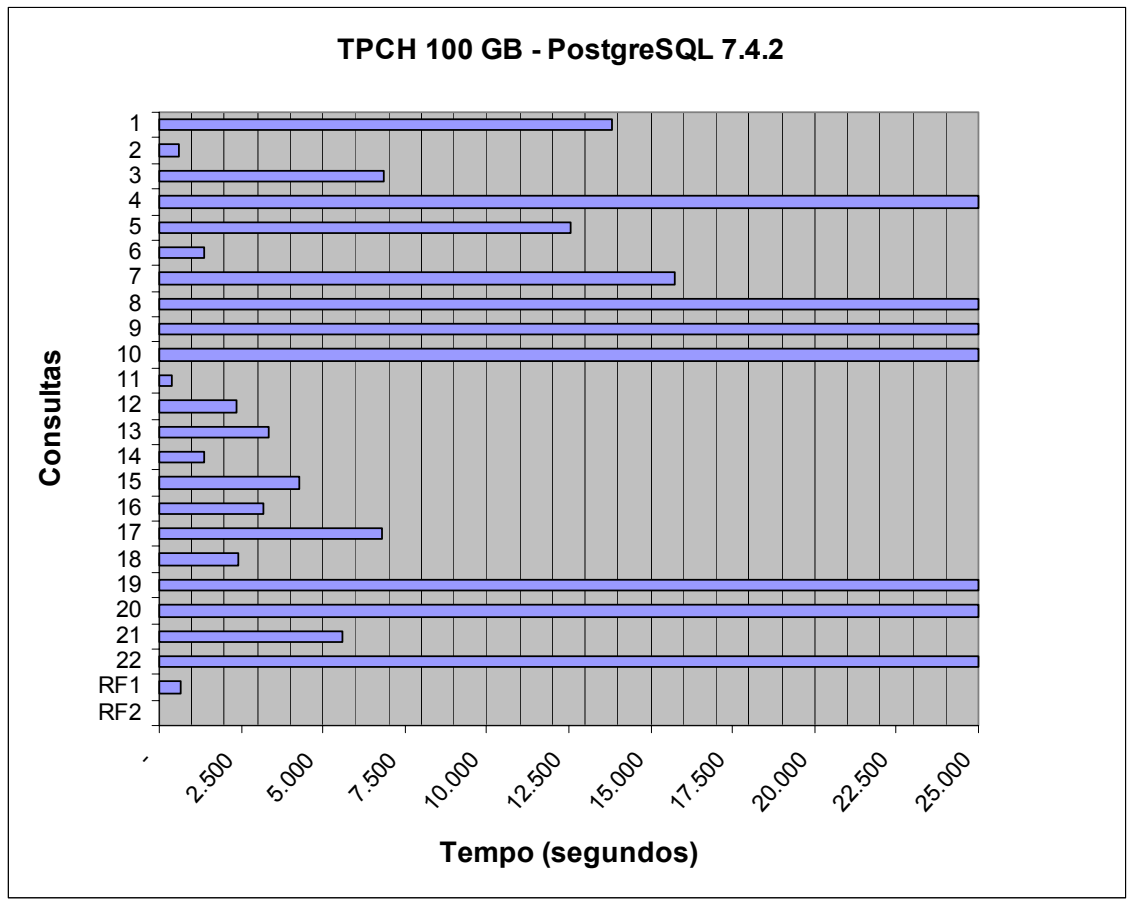

| Consulta | Tempo |
|---|---|
| 1 | 13.821 |
| 2 | 621 |
| 3 | 6.843 |
| 4 | 25.000 |
| 5 | 12.568 |
| 6 | 1.383 |
| 7 | 15.734 |
| 8 | 25.000 |
| 9 | 25.000 |
| 10 | 25.001 |
| 11 | 389 |
| 12 | 2.381 |
| 13 | 3.371 |
| 14 | 1.374 |
| 15 | 4.273 |
| 16 | 3.171 |
| 17 | 6.780 |
| 18 | 2.429 |
| 19 | 25.000 |
| 20 | 25.001 |
| 21 | 5.577 |
| 22 | 25.001 |
| RF1 | 645 |
| RF2 | 61 |

As consultas que consumiram o maior tempo mostraram algumas deficiências do PostgreSQL em comparação com outros SGBDs. Estas consultas são as de numero 4, 8, 9, 10, 19, 20 e 22. As operações realizadas foram:

- Sub-consultas dentro de outra sub-consulta;
- Operadores EXISTS e NOT EXISTS;
- Agregações quando utilizado visões *in-line*, que são sub-consultas dentro da clausula FROM;
- Seleções por data.

Muitas destas consultas realizam seleções por data, que são operações muito comuns em *data warehouse*, e SGBDs, como o Sybase, possuem índices específicos para esta finalidade.

Outro importante gargalo que encontramos no desempenho do PostgreSQL é o otimizador.

O otimizador cria o plano de execução após a reescrita da consulta, baseando-se nas estatísticas coletadas pelo DBA, portanto dois fatores são importantes: a coleta periódica das estatísticas e a reescrita das consultas.

No primeiro fator determinamos a coleta completa das estatísticas após o carregamento dos dados, gerando informações suficientes para realizar os testes.

No segundo fator a reescrita das consultas é realizada pelo SGBD, não sendo possível alterar este processo no otimizador, neste caso é necessário alterar a consulta manualmente antes de submeter ao sistema.

No inicio dos testes, na metodologia TPC-H, utilizamos as consultas originais sendo que algumas foram interrompidas através da clausula de *timeout*. Estas interrupções foram causadas porque o PostgreSQL não conseguiu gerar um plano de execução adequado para cada consulta, motivo este que permitiu somente a execução do *power test*.

Verificamos que o plano de execução não foi gerado de forma adequada por problemas na reescrita das consultas que é um dos motivadores para o desenvolvimento da nova versão do PostgreSQL.

Podemos demonstrar o problema da reescrita nas FIGURA17 e FIGURA18.

## FIGURA 17 – SQL CONSULTA Nº 19

**Consulta 19 - Original**

```
Select  sum(l_extendedprice* (1 - l_discount)) as revenue
from
  lineitem,
  part
where
(
  p_partkey = l_partkey
  and p_brand = '[BRAND1]'
  and p_container in ('SM CASE', 'SM BOX', 'SM PACK', 'SM PKG')
  and l_quantity >= [QUANTITY1] and l_quantity <= [QUANTITY1] + 10
  and p_size between 1 and 5
  and l_shipmode in ('AIR', 'AIR REG')
  and l_shipinstruct = 'DELIVER IN PERSON'
)or
(
  p_partkey = l_partkey
  and p_brand = '[BRAND2]'
  and p_container in ('MED BAG', 'MED BOX', 'MED PKG', 'MED PACK')
  and l_quantity >= [QUANTITY2] and l_quantity <= [QUANTITY2] + 10
  and p_size between 1 and 10
  and l_shipmode in ('AIR', 'AIR REG')
  and l_shipinstruct = 'DELIVER IN PERSON'
)or
...
```

**Consulta 19 - Reescrita**

```
Select   sum(l_extendedprice* (1 - l_discount)) as revenue
from
  lineitem,
  part
where
  p_partkey = l_partkey
  and l_shipmode in ('AIR', 'AIR REG')
  and l_shipinstruct = 'DELIVER IN PERSON'
  and
( (
  p_brand = '[BRAND1]'
  and p_container in ('SM CASE', 'SM BOX', 'SM PACK', 'SM PKG')
  and l_quantity >= [QUANTITY1] and l_quantity <= [QUANTITY1] +10
  and p_size between 1 and 5
) or
(
  p_brand = '[BRAND2]'
  and p_container in ('MED BAG', 'MED BOX', 'MED PKG', 'MED PACK')
  and l_quantity >= [QUANTITY2] and l_quantity <= [QUANTITY2] +10
  and p_size between 1 and 10
)or
  ...
```

Na FIGURA 17 observamos que na consulta original as cláusulas de junção são realizadas em cada operação de "OR", bem como algumas seleções. Na consulta reescrita os segmentos comuns são deixados fora da operação "OR". Este tipo de reescrita poderia ter sido realizada pelo otimizador, na FIGURA 17 a nova consulta foi desenvolvida pelo OSDL para o benchmark DBT3.

Os resultados dos planos de execução para a consulta número 19 (original e modificada) podem ser vistos na FIGURA 18.

## FIGURA 18 – PLANOS DE EXECUÇAO DA CONSULTA Nº 19

**Plano de execução original**

Aggregate  cost=672136305127.42..672136305127.43 rows=1 width=22)
  -> Nested Loop  (cost=7117.00..672136305127.15 rows=108 width=22)
      Join Filter: (...........)
      -> Seq Scan on lineitem  (cost=0.00..218010.15 rows=6001215 width=79)
      -> Materialize  (cost=7117.00..9117.00 rows=200000 width=36)
          -> Seq Scan on part  (cost=0.00..7117.00 rows=200000 width=36)
(6 rows)

**Plano de execução após reescrita**

Aggregate  cost=288750.64..288750.64 rows=1 width=22)
  -> Hash Join  (cost=7617.00..288750.37 rows=104 width=22)
      Hash Cond: ("outer".l_partkey = "inner".p_partkey)
      Join Filter: ((("inner".p_brand = 'Brand#44'::bpchar) AND (("inner".p_container = 'SM CASE'::bpchar) OR ("inner".p_container = 'SM BOX'::bpchar) OR ("inner".p_container = 'SM PACK'::bpchar) OR ("inner".p_container = 'SM PKG'::bpchar)) AND ("outer".l_quantity >= 6::numeric) AND ("outer".l_quantity <= 16::numeric) AND ("inner".p_size >= 1) AND ("inner".p_size <= 5)) OR (("inner".p_brand = 'Brand#21'::bpchar) AND (("inner".p_container = 'MED BAG'::bpchar) OR ("inner".p_container = 'MED BOX'::bpchar) OR ("inner".p_container = 'MED PKG'::bpchar) OR ("inner".p_container = 'MED PACK'::bpchar)) AND ("outer".l_quantity >= 11::numeric) AND ("outer".l_quantity <= 21::numeric) AND ("inner".p_size >= 1) AND ("inner".p_size <= 10)) OR (("inner".p_brand = 'Brand#21'::bpchar) AND (("inner".p_container = 'LG CASE'::bpchar) OR ("inner".p_container = 'LG BOX'::bpchar) OR ("inner".p_container = 'LG PACK'::bpchar) OR ("inner".p_container = 'LG PKG'::bpchar)) AND ("outer".l_quantity >= 23::numeric) AND ("outer".l_quantity <= 33::numeric) AND ("inner".p_size >= 1) AND ("inner".p_size <= 15)))
      -> Seq Scan on lineitem  (cost=0.00..263019.26 rows=219565 width=36)
          Filter: (((l_shipmode = 'AIR'::bpchar) OR (l_shipmode = 'AIR REG'::bpchar)) AND (l_shipinstruct = 'DELIVER IN PERSON'::bpchar))
      -> Hash  (cost=7117.00..7117.00 rows=200000 width=36)
          -> Seq Scan on part  (cost=0.00..7117.00 rows=200000 width=36)
(8 rows)

No plano de execução é calculado o custo da consulta de acordo com as operações a serem realizadas. O otimizador verifica qual a melhor operação bem como a melhor ordem para cada operação. As operações podem ser de agregação, junção, seleção, varredura completa entre outras.

A intenção na otimização do SGBD é que o custo seja o menor possível e no caso da consulta de número 19 somente a reescrita manual pode melhorar este custo.

Neste caso optamos por interromper as consultas com um *timeout* de 25.000 segundos como referido anteriormente.

Com a finalização por *timeout*, a análise dos planos de execução e dos textos SQL, foi possível verificar as operações em que o SGBD possui maior dificuldade.

Após ser promovida a interrupção das consultas por *timeout,* algumas foram executadas sem esta restrição. Mesmo sem esta restrição a consulta de numero 19 teve que ser interrompida, pois passou de 72 horas de execução na escala de 100GB.

Na escala de 1GB, esta mesma consulta foi interrompida após 24 horas de execução. Quando foi utilizada a consulta reescrita ela foi processada em 25,64 segundos em média, o que confirma o problema com o otimizador do PostgreSQL, mais especificamente no modulo de reescrita de consultas.

O SGBD Sybase na escala de 100 GB executou a consulta número 19 original em 971,7 segundos em média. É ilustrado no QUADRO 2 o ambiente do *benchmark* do Sybase, ambiente que é similar ao utilizado neste trabalho.

Não foi possível comparar os resultados dos testes aferidos com os divulgados pelo TPC-H, pois os testes tiveram que ser interrompidos por *timeout* ou por excederem vários dias de execução, neste último caso decidimos que é inaceitável.

A partir desta situação executamos o DBT3, pois considera a reescrita das consultas promovida pelo OSDL. Nesta condição o PostgreSQL processou a carga de trabalho.

Os resultados referentes aos testes da escala de 1GB seguem abaixo. A FIGURA 19 mostra o gráfico e os tempos de execução do teste.

## FIGURA 19 – RESULTADO OSDL DBT3 1GB

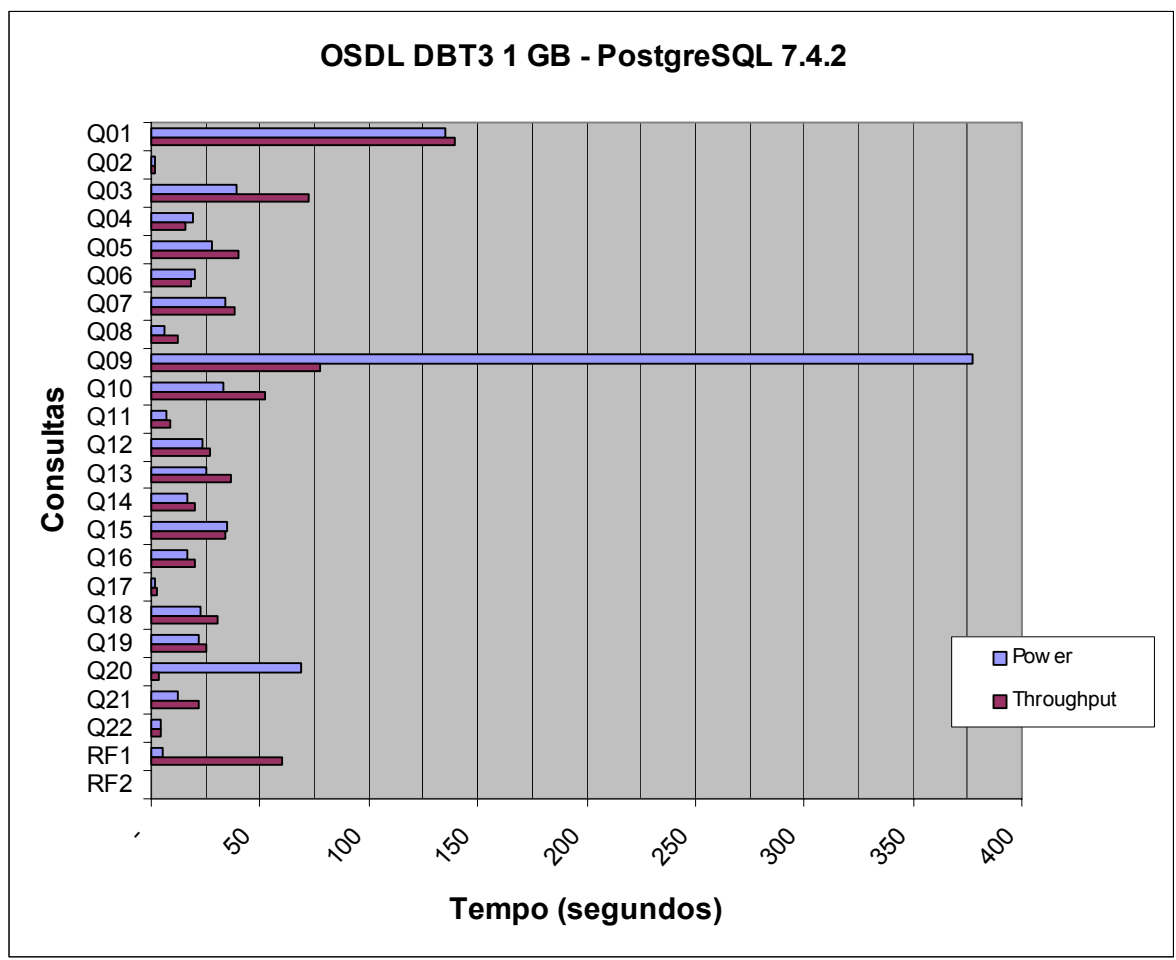

| | Power | Throughput | | Média(T1 e T2) | Maior(T1 e T2) | Menor(T1 e T2) |
|---|---|---|---|---|---|---|
| | | 1 | 2 | | | |
| Consulta | | | | | | |
| Q1 | 135,292472 | 138,626762 | 140,643522 | 139,635142 | 140,643522 | 138,626762 |
| Q2 | 1,748124 | 1,995163 | 1,089021 | 1,542092 | 1,995163 | 1,089021 |
| Q3 | 39,181041 | 102,552385 | 41,701793 | 72,127089 | 102,552385 | 41,701793 |
| Q4 | 19,531001 | 11,838140 | 19,678958 | 15,758549 | 19,678958 | 11,838140 |
| Q5 | 28,016644 | 44,981659 | 34,952009 | 39,966834 | 44,981659 | 34,952009 |
| Q6 | 20,447991 | 16,562057 | 20,366539 | 18,464298 | 20,366539 | 16,562057 |
| Q7 | 33,926596 | 36,586992 | 40,740274 | 38,663633 | 40,740274 | 36,586992 |
| Q8 | 6,214762 | 13,039518 | 10,667359 | 11,853439 | 13,039518 | 10,667359 |
| Q9 | 377,366720 | 72,558255 | 81,909059 | 77,233657 | 81,909059 | 72,558255 |
| Q10 | 33,417731 | 46,324082 | 57,518423 | 51,921253 | 57,518423 | 46,324082 |
| Q11 | 7,093706 | 10,325813 | 7,364961 | 8,845387 | 10,325813 | 7,364961 |
| Q12 | 23,916278 | 29,520291 | 25,294525 | 27,407408 | 29,520291 | 25,294525 |
| Q13 | 25,430390 | 35,260513 | 37,465099 | 36,362806 | 37,465099 | 35,260513 |
| Q14 | 16,144596 | 15,716307 | 24,112982 | 19,914645 | 24,112982 | 15,716307 |
| Q15 | 34,546518 | 31,695052 | 36,570590 | 34,132821 | 36,570590 | 31,695052 |
| Q16 | 16,759939 | 19,609934 | 21,124462 | 20,367198 | 21,124462 | 19,609934 |
| Q17 | 1,406296 | 2,217687 | 3,721933 | 2,969810 | 3,721933 | 2,217687 |
| Q18 | 22,827049 | 31,859746 | 28,715011 | 30,287379 | 31,859746 | 28,715011 |
| Q19 | 21,922508 | 21,007594 | 30,282406 | 25,645000 | 30,282406 | 21,007594 |
| Q20 | 68,575981 | 3,820619 | 3,281648 | 3,551134 | 3,820619 | 3,281648 |
| Q21 | 12,328835 | 27,867894 | 15,774713 | 21,821304 | 27,867894 | 15,774713 |
| Q22 | 4,788507 | 4,962036 | 4,075970 | 4,519003 | 4,962036 | 4,075970 |
| RF1 | 5,467944 | 51,683042 | 68,121353 | 59,902198 | 68,121353 | 51,683042 |
| RF2 | 0,012504 | 0,026950 | 0,015701 | 0,021326 | 0,026950 | 0,015701 |

As métricas geradas pelo *benchmark* OSDL DBT3 foram:

- Power@size = 332,35

- Throughput@size = 224,85

- Composite = 273,37

Estes resultados somente são comparáveis aos divulgados pelo OSDL. Neste caso somente encontramos testes do próprio PostgreSQL, não sendo conclusivo para comparação com outro SGBD.

O QUADRO 11 apresenta um resumo da execução dos *benchmarks* deste trabalho.

QUADRO 11 – RESULTADOS DOS *BENCHMARKS*

| Benchmark | Escala (GB) | Timeout | | | Sem Timeout | |
| | | Situação | Timeout por consulta (seg) | Tempo final (seg) | Situação | Tempo final |
|---|---|---|---|---|---|---|
| TPC-H | 100 | Executado (*) | 25.000 | 256.424 | Interrompido | > 72Hr |
| TPC-H | 1 | - | - | - | Interrompido | > 24Hr |
| DBT3 | 1 | - | - | - | Executado | 2.482 seg |

(*) somente power test

# 9 CONCLUSÃO E TRABALHOS FUTUROS

Neste trabalho buscamos realizar *benchmarks* que atestem o desempenho de uma plataforma de baixo custo utilizando SGBD PostgreSQL como repositório de um data warehouse e sistema operacional GNU/Linux, apontando as deficiências e verificando se mesmo com estas deficiências a plataforma é viável.

Verificamos também estruturas que venham a aumentar o desempenho do PostgreSQL buscando enriquecer as possibilidades de melhorias neste SGBD.

Pelos resultados aferidos ficou demonstrado que a versão 7.4.x do PostgreSQL, não conseguiu executar satisfatoriamente o *benchmark* TPC-H que representa uma carga de trabalho do mundo real, tendo sido necessário executar o *benchmark* DBT3.

Logo, a expectativa de viabilidade de uma plataforma de baixo custo em um data *warehouse,* com o ambiente utilizado neste estudo e sem a imposição de restrições, não foi plenamente alcançada não sendo possível comparar com as estruturas publicadas pelo TPC.

Conclui-se que a atual versão do PostgreSQL é inviável para *Data Warehouse* com a estratégia de reescrita de consultas que é empregada no otimizador. Esta falha no otimizador acarreta um tempo de execução exponencialmente maior comparado com outros resultados publicados pelo TPC, como demonstrado no estudo da consulta de número 19.

É importante mencionar a considerável melhoria no desempenho do PostgreSQL da versão 7.3.4 para 7.4.2 com a implementação de algumas características para operações de agregação. Podemos, portanto, esperar que num futuro próximo a diferença em comparação com Sybase e MS SQL Server diminua, considerando as implementações que estamos sugerindo e a evolução natural no desenvolvimento do SGBD.

Esperamos que nas próximas versões do PostgreSQL seja possível utilizá-lo em um projeto de *data warehouse* em qualquer escala, se os problemas no otimizador forem corrigidos e as estruturas que sugerimos forem implementadas. A correção dos problemas no otimizador é considerada crucial mesmo com a implementação das estruturas que apontamos neste trabalho.

Como trabalhos futuros podemos sugerir a implementação, no SGBD PostgreSQL, das estruturas que melhoram seu desempenho. Estas estruturas seriam:

- Estratégias de armazenamento como o PAX, proposto por AILAMAKI et al (2001). Esta estratégia pode diminuir as operações de E/S como descrito pelo autor;

- Índice de bitmap como o *encoded bitmap,* proposto por WU e BUCHMANN (1998). Este tipo de índice pode aumentar consideravelmente o desempenho de consultas com agregação;

- Paralelismo *intra-query* descrito por OMIECINSKI (1995). O paralelismo pode melhorar o desempenho das sub-consultas e visões *in-line* que observamos terem baixo desempenho no PostgreSQL, bem como se beneficiar de ambientes multiprocessados;

- Execução dos testes em servidores dispostos em *cluster* para verificar se haverá aumento de desempenho, pois neste estudo está sendo utilizado equipamento multiprocessado. Servidores em *cluster* são uma alternativa quando é necessário aumentar o poder de processamento de um sistema e possuem custo baixo em comparação com sistemas multiprocessados;

- Desenvolvimento de uma metodologia correlata ao TPC-H e OSDL DBT3 para tirar proveito das características de orientação a objeto que SGBDs como o PostgreSQL possui.

Outros trabalhos poderiam ser desenvolvidos, seguindo o modelo de software livre, para os demais componentes do ambiente *data warehouse,* como a implementação de ferramentas, tais como:

- *On-Line Analytical Processing* (OLAP);

- *Executive Information Systems* (EIS);

- *Extract, Transoform and Load* (ETL) que se beneficiaria dos trabalhos desenvolvidos por KAVALCO (2001) e RODACKI (2000).

**ANEXOS**

# 1 CODIGO DA FUNÇÃO RF1

```
import java.io.*;
import java.sql.*;
import java.util.*;
public class Insert
{
    public static void main(String[] args)
    {
        System.out.println("Inicio RF1");
        int times=0;
        try{
            try{
                Class.forName("org.postgresql.Driver");
            } catch (Exception e) {
                System.out.println(e.getMessage());
            }
            //Conecta-se com o driver
            String url = "jdbc:postgresql:db_eduardo";
            String user = "eduardo";
            String password = "bmTPCH2004";
            Connection con = DriverManager.getConnection(url,user,password);

            //Cria um statement para enviar instrucoes para o driver
            Statement stmt = con.createStatement();

            BufferedReader bfArqOrders = new BufferedReader(new
FileReader("orders.tbl.u1"));
            BufferedReader bfArqLineitem = new BufferedReader(new
FileReader("lineitem.tbl.u1"));
            String linhaOrders= bfArqOrders.readLine();
            String linhaLineitem= bfArqLineitem.readLine();

            Random rdm = new Random();
            while (linhaOrders != null){ // le todas as linhas
                StringTokenizer stOrders=new StringTokenizer(linhaOrders,"|");
                    String o_orderkey = stOrders.nextToken();
                    String o_custkey = stOrders.nextToken();
                    String o_orderstatus = stOrders.nextToken();
                    String o_totalprice = stOrders.nextToken();
                    String o_orderdate = stOrders.nextToken();
                    String o_orderpriority = stOrders.nextToken();
                    String o_clerk = stOrders.nextToken();
                    String o_shippriority = stOrders.nextToken();
                    String o_comment = stOrders.nextToken();
                    String sql = "insert into orders2
values("+o_orderkey+","+o_custkey+",'"+o_orderstatus+"',"+o_totalprice+",'"+o_orderdat
e+"','"+o_orderpriority+"','"+o_clerk+"',"+o_shippriority+",'"+o_comment+"');";

                    System.out.println(sql);
                    times = rdm.nextInt(7);
                stmt.executeUpdate(sql);
```

```
                do{  // randomico de 7x
                    if (linhaLineitem != null){ // le todas as linhas
                        StringTokenizer stLineitem=new
StringTokenizer(linhaLineitem,"|");
                        String l_orderkey = stLineitem.nextToken();
                        String l_partkey = stLineitem.nextToken();
                        String l_suppkey = stLineitem.nextToken();
                        String l_linenumber = stLineitem.nextToken();
                        String l_quantity = stLineitem.nextToken();
                        String l_extendedprice = stLineitem.nextToken();
                        String l_discount = stLineitem.nextToken();
                        String l_tax = stLineitem.nextToken();
                        String l_returnflag = stLineitem.nextToken();
                        String l_linestatus = stLineitem.nextToken();
                        String l_shipdate = stLineitem.nextToken();
                        String l_commitdate = stLineitem.nextToken();
                        String l_receiptdate = stLineitem.nextToken();
                        String l_shipinstruct = stLineitem.nextToken();
                        String l_shipmode = stLineitem.nextToken();
                        String l_comment = stLineitem.nextToken();
                        String sql2 = "insert into lineitem2
values("+l_orderkey+","+l_partkey+","+l_suppkey+","+l_linenumber+","+l_quantity+","+l
_extendedprice+","+l_discount+","+l_tax+","'+l_returnflag+"','"+l_linestatus+"','"+l_ship
date+"','"+l_commitdate+"','"+l_receiptdate+"','"+l_shipinstruct+"','"+l_shipmode+"','"+l_
comment+"');";
                        stmt.executeUpdate(sql2);
                    }
                    times--;
                    linhaLineitem= bfArqLineitem.readLine();
                } while (times > 0);
                linhaOrders= bfArqOrders.readLine();
            }

            bfArqOrders.close();
            con.commit();
            con.close();
            System.out.println("Termino RF1");
        }catch(Exception io){
            System.out.println(io.getMessage());
        }
    }//main
}//class
```

## 2 CODIGO DA FUNÇÃO RF2

```
import java.io.*;
import java.sql.*;
public class Delete
{
        public static void main(String[] args)
        {
                System.out.println("Inicio RF2");
                try{
                        try{
                                Class.forName("org.postgresql.Driver");
                        } catch (Exception e) {
                                System.out.println(e.getMessage());
                        }
                        //Conecta-se com o driver
                        String url = "jdbc:postgresql:db_eduardo";
                        String user = "eduardo";
                        String password = "bmTPCH2004";
                        Connection con = DriverManager.getConnection(url,user,password);
                        //Cria um statement para enviar instrucoes para o driver
                        Statement stmt = con.createStatement();

                        BufferedReader bfArq = new BufferedReader(new
FileReader(args[0]));
                        String linha= bfArq.readLine();
                        while (linha != null){// le todas as linhas
                                stmt.executeUpdate("delete from orders2 where o_orderkey
="+ linha +";");
                                stmt.executeUpdate("delete from lineitem2 where l_orderkey
="+ linha +";");
                                linha= bfArq.readLine();
                        }

                        bfArq.close();
                        con.commit();
                        con.close();
                        System.out.println("Termino RF2");
                }catch(Exception io){
                        System.out.println(io.getMessage());
                }
        }//main
}//class
```

# 3 CONSULTAS SUBMETIDAS AO SGBD (*POWER TEST*)

```
-- using 1782942327 as a seed to the RNG
\timing
\! echo "---Inicio teste---" >> log.txt
\! echo |date +%H:%M:%S >> log.txt

\! echo "---q14 ini---" >> log.txt
\! echo |date "+%H-%M-%S" >> log.txt

-- @(#)14.sql     2.1.8.1
-- TPC-H/TPC-R Promotion Effect Query (Q14)
-- Functional Query Definition
-- Approved February 1998

select
      100.00 * sum(case
            when p_type like 'PROMO%'
                  then l_extendedprice * (1 - l_discount)
            else 0
      end) / sum(l_extendedprice * (1 - l_discount)) as promo_revenue
from
      lineitem,
      part
where
      l_partkey = p_partkey
      and l_shipdate >= date '1998-01-01'
      and l_shipdate < date '1998-01-01' + interval '1 month';
\! echo "---q2 ini---" >> log.txt
\! echo |date "+%H-%M-%S" >> log.txt

-- @(#)2.sql     2.1.8.2
-- TPC-H/TPC-R Minimum Cost Supplier Query (Q2)
-- Functional Query Definition
-- Approved February 1998

select
      s_acctbal,
      s_name,
      n_name,
      p_partkey,
      p_mfgr,
      s_address,
      s_phone,
      s_comment
from
      part,
      supplier,
      partsupp,
      nation,
      region
where
      p_partkey = ps_partkey
      and s_suppkey = ps_suppkey
      and p_size = 23
```

```
        and p_type like '%STEEL'
        and s_nationkey = n_nationkey
        and n_regionkey = r_regionkey
        and r_name = 'AFRICA'
        and ps_supplycost = (
                select
                        min(ps_supplycost)
                from
                        partsupp,
                        supplier,
                        nation,
                        region
                where
                        p_partkey = ps_partkey
                        and s_suppkey = ps_suppkey
                        and s_nationkey = n_nationkey
                        and n_regionkey = r_regionkey
                        and r_name = 'AFRICA'
        )
order by
        s_acctbal desc,
        n_name,
        s_name,
        p_partkey;
\! echo "---q9 ini---" >> log.txt
\! echo |date "+%H-%M-%S" >> log.txt

-- @(#)9.sql    2.1.8.1
-- TPC-H/TPC-R Product Type Profit Measure Query (Q9)
-- Functional Query Definition
-- Approved February 1998

select
        nation,
        o_year,
        sum(amount) as sum_profit
from
        (
                select
                        n_name as nation,
                        extract(year from o_orderdate) as o_year,
                        l_extendedprice * (1 - l_discount) - ps_supplycost *
l_quantity as amount
                from
                        part,
                        supplier,
                        lineitem,
                        partsupp,
                        orders,
                        nation
                where
                        s_suppkey = l_suppkey
                        and ps_suppkey = l_suppkey
                        and ps_partkey = l_partkey
                        and p_partkey = l_partkey
                        and o_orderkey = l_orderkey
                        and s_nationkey = n_nationkey
                        and p_name like '%tan%'
```

```
      ) as profit
group by
      nation,
      o_year
order by
      nation,
      o_year desc;
\! echo "---q20 ini---" >> log.txt
\! echo |date "+%H-%M-%S" >> log.txt

-- @(#)20.sql     2.1.8.1
-- TPC-H/TPC-R Potential Part Promotion Query (Q20)
-- Function Query Definition
-- Approved February 1998

select
      s_name,
      s_address
from
      supplier,
      nation
where
      s_suppkey in (
            select
                  ps_suppkey
            from
                  partsupp
            where
                  ps_partkey in (
                        select
                              p_partkey
                        from
                              part
                        where
                              p_name like 'royal%'
                  )
                  and ps_availqty > (
                        select
                              0.5 * sum(l_quantity)
                        from
                              lineitem
                        where
                              l_partkey = ps_partkey
                              and l_suppkey = ps_suppkey
                              and l_shipdate >= date '1997-01-01'
                              and l_shipdate < date '1997-01-01' + interval
'1 year'
                  )
      )
      and s_nationkey = n_nationkey
      and n_name = 'EGYPT'
order by
      s_name;

\! echo "---q6 ini---" >> log.txt
\! echo |date "+%H-%M-%S" >> log.txt
```

```
-- @(#)6.sql       2.1.8.1
-- TPC-H/TPC-R Forecasting Revenue Change Query (Q6)
-- Functional Query Definition
-- Approved February 1998

select
      sum(l_extendedprice * l_discount) as revenue
from
      lineitem
where
      l_shipdate >= date '1995-01-01'
      and l_shipdate < date '1995-01-01' + interval '1 year'
      and l_discount between 0.02 - 0.01 and 0.02 + 0.01
      and l_quantity < 25;
\! echo "---q17 ini---" >> log.txt
\! echo |date "+%H-%M-%S" >> log.txt

-- @(#)17.sql      2.1.8.1
-- TPC-H/TPC-R Small-Quantity-Order Revenue Query (Q17)
-- Functional Query Definition
-- Approved February 1998

select
      sum(l_extendedprice) / 7.0 as avg_yearly
from
      lineitem,
      part
where
      p_partkey = l_partkey
      and p_brand = 'Brand#13'
      and p_container = 'MED PACK'
      and l_quantity < (
            select
                  0.2 * avg(l_quantity)
            from
                  lineitem
            where
                  l_partkey = p_partkey
      );
\! echo "---q18 ini---" >> log.txt
\! echo |date "+%H-%M-%S" >> log.txt

-- @(#)18.sql      2.1.8.1
-- TPC-H/TPC-R Large Volume Customer Query (Q18)
-- Function Query Definition
-- Approved February 1998

select
      c_name,
      c_custkey,
      o_orderkey,
      o_orderdate,
      o_totalprice,
      sum(l_quantity)
from
      customer,
```

```
        orders,
        lineitem
where
        o_orderkey in (
                select
                        l_orderkey
                from
                        lineitem
                group by
                        l_orderkey having
                                sum(l_quantity) > 314
        )
        and c_custkey = o_custkey
        and o_orderkey = l_orderkey
group by
        c_name,
        c_custkey,
        o_orderkey,
        o_orderdate,
        o_totalprice
order by
        o_totalprice desc,
        o_orderdate;
\! echo "---q8 ini---" >> log.txt
\! echo |date "+%H-%M-%S" >> log.txt

-- @(#)8.sql      2.1.8.1
-- TPC-H/TPC-R National Market Share Query (Q8)
-- Functional Query Definition
-- Approved February 1998

select
        o_year,
        sum(case
                when nation = 'JORDAN' then volume
                else 0
        end) / sum(volume) as mkt_share
from
        (
                select
                        extract(year from o_orderdate) as o_year,
                        l_extendedprice * (1 - l_discount) as volume,
                        n2.n_name as nation
                from
                        part,
                        supplier,
                        lineitem,
                        orders,
                        customer,
                        nation n1,
                        nation n2,
                        region
                where
                        p_partkey = l_partkey
                        and s_suppkey = l_suppkey
                        and l_orderkey = o_orderkey
                        and o_custkey = c_custkey
                        and c_nationkey = n1.n_nationkey
```

```
                and n1.n_regionkey = r_regionkey
                and r_name = 'MIDDLE EAST'
                and s_nationkey = n2.n_nationkey
                and o_orderdate between date '1995-01-01' and date '1996-
12-31'
                and p_type = 'PROMO PLATED BRASS'
    ) as all_nations
group by
        o_year
order by
        o_year;
\! echo "---q21 ini---" >> log.txt
\! echo |date "+%H-%M-%S" >> log.txt

-- @(#)21.sql    2.1.8.1
-- TPC-H/TPC-R Suppliers Who Kept Orders Waiting Query (Q21)
-- Functional Query Definition
-- Approved February 1998

select
        s_name,
        count(*) as numwait
from
        supplier,
        lineitem l1,
        orders,
        nation
where
        s_suppkey = l1.l_suppkey
        and o_orderkey = l1.l_orderkey
        and o_orderstatus = 'F'
        and l1.l_receiptdate > l1.l_commitdate
        and exists (
            select
                    *
            from
                    lineitem l2
            where
                    l2.l_orderkey = l1.l_orderkey
                    and l2.l_suppkey <> l1.l_suppkey
        )
        and not exists (
            select
                    *
            from
                    lineitem l3
            where
                    l3.l_orderkey = l1.l_orderkey
                    and l3.l_suppkey <> l1.l_suppkey
                    and l3.l_receiptdate > l3.l_commitdate
        )
        and s_nationkey = n_nationkey
        and n_name = 'RUSSIA'
group by
        s_name
order by
        numwait desc,
        s_name;
```

```
\! echo "---q13 ini---" >> log.txt
\! echo |date "+%H-%M-%S" >> log.txt

-- @(#)13.sql    2.1.8.1
-- TPC-H/TPC-R Customer Distribution Query (Q13)
-- Functional Query Definition
-- Approved February 1998

select
      c_count,
      count(*) as custdist
from
      (
            select
                  c_custkey,
                  count(o_orderkey)
            from
                  customer left outer join orders on
                        c_custkey = o_custkey
                        and o_comment not like '%special%deposits%'
            group by
                  c_custkey
      ) as c_orders (c_custkey, c_count)
group by
      c_count
order by
      custdist desc,
      c_count desc;
\! echo "---q3 ini---" >> log.txt
\! echo |date "+%H-%M-%S" >> log.txt

-- @(#)3.sql    2.1.8.1
-- TPC-H/TPC-R Shipping Priority Query (Q3)
-- Functional Query Definition
-- Approved February 1998

select
      l_orderkey,
      sum(l_extendedprice * (1 - l_discount)) as revenue,
      o_orderdate,
      o_shippriority
from
      customer,
      orders,
      lineitem
where
      c_mktsegment = 'FURNITURE'
      and c_custkey = o_custkey
      and l_orderkey = o_orderkey
      and o_orderdate < date '1995-03-01'
      and l_shipdate > date '1995-03-01'
group by
      l_orderkey,
      o_orderdate,
      o_shippriority
order by
      revenue desc,
```

```
      o_orderdate;
\! echo "---q22 ini---" >> log.txt
\! echo |date "+%H-%M-%S" >> log.txt

-- @(#)22.sql     2.1.8.1
-- TPC-H/TPC-R Global Sales Opportunity Query (Q22)
-- Functional Query Definition
-- Approved February 1998

select
      cntrycode,
      count(*) as numcust,
      sum(c_acctbal) as totacctbal
from
      (
            select
                  substr(c_phone,1,2) as cntrycode,
                  c_acctbal
            from
                  customer
            where
                  substr(c_phone,1,2) in
                        ('11', '14', '25', '15', '21', '17', '20')
                  and c_acctbal > (
                        select
                              avg(c_acctbal)
                        from
                              customer
                        where
                              c_acctbal > 0.00
                              and substr(c_phone,1,2) in
                                    ('11', '14', '25', '15', '21', '17',
'20')
                  )
                  and not exists (
                        select
                              *
                        from
                              orders
                        where
                              o_custkey = c_custkey
                  )
      ) as custsale
group by
      cntrycode
order by
      cntrycode;

\! echo "---q16 ini---" >> log.txt
\! echo |date "+%H-%M-%S" >> log.txt

-- @(#)16.sql     2.1.8.1
-- TPC-H/TPC-R Parts/Supplier Relationship Query (Q16)
-- Functional Query Definition
-- Approved February 1998

select
```

```
      p_brand,
      p_type,
      p_size,
      count(distinct ps_suppkey) as supplier_cnt
from
      partsupp,
      part
where
      p_partkey = ps_partkey
      and p_brand <> 'Brand#41'
      and p_type not like 'LARGE BURNISHED%'
      and p_size in (26, 48, 45, 7, 41, 46, 31, 17)
      and ps_suppkey not in (
            select
                  s_suppkey
            from
                  supplier
            where
                  s_comment like '%Customer%Complaints%'
      )
group by
      p_brand,
      p_type,
      p_size
order by
      supplier_cnt desc,
      p_brand,
      p_type,
      p_size;
\! echo "---q4 ini---" >> log.txt
\! echo |date "+%H-%M-%S" >> log.txt

-- @(#)4.sql      2.1.8.1
-- TPC-H/TPC-R Order Priority Checking Query (Q4)
-- Functional Query Definition
-- Approved February 1998

select
      o_orderpriority,
      count(*) as order_count
from
      orders
where
      o_orderdate >= date '1993-03-01'
      and o_orderdate < date '1993-03-01' + interval '3 month'
      and exists (
            select
                  *
            from
                  lineitem
            where
                  l_orderkey = o_orderkey
                  and l_commitdate < l_receiptdate
      )
group by
      o_orderpriority
order by
```

```
      o_orderpriority;
\! echo "---q11 ini---" >> log.txt
\! echo |date "+%H-%M-%S" >> log.txt

-- @(#)11.sql     2.1.8.1
-- TPC-H/TPC-R Important Stock Identification Query (Q11)
-- Functional Query Definition
-- Approved February 1998

select
      ps_partkey,
      sum(ps_supplycost * ps_availqty) as value
from
      partsupp,
      supplier,
      nation
where
      ps_suppkey = s_suppkey
      and s_nationkey = n_nationkey
      and n_name = 'BRAZIL'
group by
      ps_partkey having
            sum(ps_supplycost * ps_availqty) > (
                  select
                        sum(ps_supplycost * ps_availqty) * 0.0001000000
                  from
                        partsupp,
                        supplier,
                        nation
                  where
                        ps_suppkey = s_suppkey
                        and s_nationkey = n_nationkey
                        and n_name = 'BRAZIL'
            )
order by
      value desc;
\! echo "---q15 ini---" >> log.txt
\! echo |date "+%H-%M-%S" >> log.txt

-- @(#)15.sql     2.1.8.1
-- TPC-H/TPC-R Top Supplier Query (Q15)
-- Functional Query Definition
-- Approved February 1998

create view revenue0 (supplier_no, total_revenue) as
      select
            l_suppkey,
            sum(l_extendedprice * (1 - l_discount))
      from
            lineitem
      where
            l_shipdate >= date '1994-08-01'
            and l_shipdate < date '1994-08-01' + interval '3 month'
      group by
            l_suppkey;

select
```

```
      s_suppkey,
      s_name,
      s_address,
      s_phone,
      total_revenue
from
      supplier,
      revenue0
where
      s_suppkey = supplier_no
      and total_revenue = (
            select
                  max(total_revenue)
            from
                  revenue0
      )
order by
      s_suppkey;

drop view revenue0;
\! echo "---q1 ini---" >> log.txt
\! echo |date "+%H-%M-%S" >> log.txt

-- @(#)1.sql      2.1.8.1
-- TPC-H/TPC-R Pricing Summary Report Query (Q1)
-- Functional Query Definition
-- Approved February 1998

select
      l_returnflag,
      l_linestatus,
      sum(l_quantity) as sum_qty,
      sum(l_extendedprice) as sum_base_price,
      sum(l_extendedprice * (1 - l_discount)) as sum_disc_price,
      sum(l_extendedprice * (1 - l_discount) * (1 + l_tax)) as sum_charge,
      avg(l_quantity) as avg_qty,
      avg(l_extendedprice) as avg_price,
      avg(l_discount) as avg_disc,
      count(*) as count_order
from
      lineitem
where
      l_shipdate <= date '1998-12-01' - interval '64 day'
group by
      l_returnflag,
      l_linestatus
order by
      l_returnflag,
      l_linestatus;
\! echo "---q10 ini---" >> log.txt
\! echo |date "+%H-%M-%S" >> log.txt

-- @(#)10.sql     2.1.8.1
-- TPC-H/TPC-R Returned Item Reporting Query (Q10)
-- Functional Query Definition
-- Approved February 1998
```

```
select
        c_custkey,
        c_name,
        sum(l_extendedprice * (1 - l_discount)) as revenue,
        c_acctbal,
        n_name,
        c_address,
        c_phone,
        c_comment
from
        customer,
        orders,
        lineitem,
        nation
where
        c_custkey = o_custkey
        and l_orderkey = o_orderkey
        and o_orderdate >= date '1993-12-01'
        and o_orderdate < date '1993-12-01' + interval '3 month'
        and l_returnflag = 'R'
        and c_nationkey = n_nationkey
group by
        c_custkey,
        c_name,
        c_acctbal,
        c_phone,
        n_name,
        c_address,
        c_comment
order by
        revenue desc;
\! echo "---q19 ini---" >> log.txt
\! echo "+%H-%M-%S" >> log.txt
-- @(#)19.sql    2.1.8.1
-- TPC-H/TPC-R Discounted Revenue Query (Q19)
-- Functional Query Definition
-- Approved February 1998
START TRANSACTION;

select
        sum(l_extendedprice* (1 - l_discount)) as revenue
from
        lineitem,
        part
where
        p_partkey = l_partkey
        and l_shipmode in ('AIR', 'AIR REG')
        and l_shipinstruct = 'DELIVER IN PERSON'
        and
        (
          (
                p_brand = 'Brand#34'
                and p_container in ('SM CASE', 'SM BOX', 'SM PACK', 'SM PKG')
                and l_quantity >= 9 and l_quantity <= 9+10
                and p_size between 1 and 5
          )
          or
```

```
      (
              p_brand = 'Brand#51'
              and p_container in ('MED BAG', 'MED BOX', 'MED PKG', 'MED
PACK')
              and l_quantity >= 13 and l_quantity <= 13+10
              and p_size between 1 and 10
          )
          or
          (
              p_brand = 'Brand#14'
              and p_container in ('LG CASE', 'LG BOX', 'LG PACK', 'LG PKG')
              and l_quantity >= 23 and l_quantity <= 23+10
              and p_size between 1 and 15
          )
       );
COMMIT;

\! echo "---q5 ini---" >> log.txt
\! echo |date "+%H-%M-%S" >> log.txt

-- @(#)5.sql      2.1.8.1
-- TPC-H/TPC-R Local Supplier Volume Query (Q5)
-- Functional Query Definition
-- Approved February 1998

select
      n_name,
      sum(l_extendedprice * (1 - l_discount)) as revenue
from
      customer,
      orders,
      lineitem,
      supplier,
      nation,
      region
where
      c_custkey = o_custkey
      and l_orderkey = o_orderkey
      and l_suppkey = s_suppkey
      and c_nationkey = s_nationkey
      and s_nationkey = n_nationkey
      and n_regionkey = r_regionkey
      and r_name = 'AFRICA'
      and o_orderdate >= date '1995-01-01'
      and o_orderdate < date '1995-01-01' + interval '1 year'
group by
      n_name
order by
      revenue desc;
\! echo "---q7 ini---" >> log.txt
\! echo |date "+%H-%M-%S" >> log.txt

-- @(#)7.sql      2.1.8.1
-- TPC-H/TPC-R Volume Shipping Query (Q7)
-- Functional Query Definition
```

```
-- Approved February 1998

select
        supp_nation,
        cust_nation,
        l_year,
        sum(volume) as revenue
from
        (
                select
                        n1.n_name as supp_nation,
                        n2.n_name as cust_nation,
                        extract(year from l_shipdate) as l_year,
                        l_extendedprice * (1 - l_discount) as volume
                from
                        supplier,
                        lineitem,
                        orders,
                        customer,
                        nation n1,
                        nation n2
                where
                        s_suppkey = l_suppkey
                        and o_orderkey = l_orderkey
                        and c_custkey = o_custkey
                        and s_nationkey = n1.n_nationkey
                        and c_nationkey = n2.n_nationkey
                        and (
                                (n1.n_name = 'ALGERIA' and n2.n_name = 'JORDAN')
                                or (n1.n_name = 'JORDAN' and n2.n_name = 'ALGERIA')
                        )
                        and l_shipdate between date '1995-01-01' and date '1996-
12-31'
        ) as shipping
group by
        supp_nation,
        cust_nation,
        l_year
order by
        supp_nation,
        cust_nation,
        l_year;
\! echo "---q12 ini---" >> log.txt
\! echo |date "+%H-%M-%S" >> log.txt

-- @(#)12.sql     2.1.8.1
-- TPC-H/TPC-R Shipping Modes and Order Priority Query (Q12)
-- Functional Query Definition
-- Approved February 1998

select
        l_shipmode,
        sum(case
                when o_orderpriority = '1-URGENT'
                        or o_orderpriority = '2-HIGH'
                        then 1
                else 0
```

```
        end) as high_line_count,
        sum(case
                when o_orderpriority <> '1-URGENT'
                     and o_orderpriority <> '2-HIGH'
                     then 1
                else 0
        end) as low_line_count
from
        orders,
        lineitem
where
        o_orderkey = l_orderkey
        and l_shipmode in ('FOB', 'AIR')
        and l_commitdate < l_receiptdate
        and l_shipdate < l_commitdate
        and l_receiptdate >= date '1997-01-01'
        and l_receiptdate < date '1997-01-01' + interval '1 year'
group by
        l_shipmode
order by
        l_shipmode;
\! echo "---Fim teste---" >> log.txt
\! echo |date +%H:%M:%S >> log.txt
```